\documentclass[a4paper,12pt]{article}
\usepackage{epsfig}
\usepackage{amssymb}
\usepackage{cite}
\usepackage{axodraw}

\newcommand{\hd}{{\bf h_d}}
\newcommand{\hu}{{\bf h_u}}
\newcommand{\hhu}{{\bf \hat{h}_u}}
\newcommand{\hhd}{{\bf \hat{h}_d}}
\newcommand{\Eg}{{\bf \hat{E}_g}}
\newcommand{\Eu}{{\bf \hat{E}_u}}

\newcommand{\asos}{{\bf 1}}
\newcommand{\ULQ}{{\bf {\cal U}_L^Q}}
\newcommand{\URu}{{\bf {\cal U}_R^u}}
\newcommand{\URd}{{\bf {\cal U}_R^d}}
\newcommand{\Vckm}{{\bf {V}}}
\newcommand{\Md}{{\bf \hat{M}_d}}
\newcommand{\Mu}{{\bf \hat{M}_u}} 
\newcommand{\Rh}{{\bf \hat{R}}}
\newcommand{\Ro}{{\bf {R}}}
\newcommand{\Egt}{{\bf \tilde{E}_g}}
\newcommand{\Eut}{{\bf \tilde{E}_u}}
\newcommand{\Egb}{{\bf {E}_g}}
\newcommand{\Eub}{{\bf {E}_u}}
\newcommand{\DmK}{{\Delta M_K}}
\newcommand{\DmBd}{\Delta M_{B_d}}
\newcommand{\DmBs}{\Delta M_{B_s}}
\newcommand{\DmBq}{\Delta M_{B_q}}
\newcommand{\ek}{{\epsilon_K}}
\newcommand{\epe}{{\frac{\epsilon'}{\epsilon}}}
\newcommand{\bra}[1]{\langle #1|}
\newcommand{\ket}[1]{|#1\rangle}

\begin{document}
\tolerance=100000

\begin{flushright}
FERMILAB-Pub-02/226-T\\[-1mm]
MC-TH/2002-05\\[-1mm]
hep-ph/0209306\\[-1mm]
September 2002 
\end{flushright}

\bigskip

\begin{center}
{\Large \bf Resummed Effective Lagrangian for Higgs-Mediated }\\[0.3cm] 
{\Large \bf FCNC Interactions in the CP-Violating MSSM}\\[1.7cm]

{{\large Athanasios Dedes}$^{\, a}$ 
{\large and Apostolos Pilaftsis}$^{\, b,c}$ }\\[0.5cm]
{\it $^a$Physikalisches Institut der Universit\"at Bonn,  Nu\ss allee 12, 
 D-53115 Bonn, Germany } \\[3mm]
{\it $^b$Fermilab, P.O. Box 500, Batavia IL 60510, U.S.A.}\\[3mm]
{\it $^c$Department of Physics and Astronomy, University of Manchester,\\ 
Manchester M13 9PL, United Kingdom }
\end{center}

\vspace*{0.8cm}\centerline{\bf ABSTRACT}   
\vspace{0.1cm}\noindent{\small    
We derive the general resummed effective Lagrangian for Higgs-mediated
flavour-changing  neutral-current  (FCNC) interactions  in the Minimal
Supersymmetric Standard  Model (MSSM), without resorting to particular
assumptions that rely on the squark-mass or the quark-Yukawa structure
of the theory.  In our  derivation we also  include the possibility of
explicit CP  violation  through the Cabibbo--Kobayashi--Maskawa mixing
matrix and soft supersymmetry-breaking  mass terms.  The advantages of
our resummed  FCNC  effective  Lagrangian are  explicitly demonstrated
within the  context of  phenomenologically  motivated scenarios.    We
obtain new testable predictions in the large $\tan\beta$ regime of the
theory for CP-conserving and  CP-violating observables related  to the
$K$- and $B$-meson  systems,  such as $\Delta  M_{K,B}$, $\epsilon_K$,
$\epsilon'/\epsilon$,  ${\cal  B}(B_{d,s} \to \ell^+\ell^-)$ and their
associated leptonic CP  asymmetries.   Finally, based on our  resummed
FCNC effective Lagrangian, we can  identify configurations in the soft
supersymmetry-breaking   parameter space, for  which    a  kind of   a
Glashow--Iliopoulos--Maiani-cancellation mechanism  becomes  operative
and hence all Higgs-mediated, $\tan\beta$-enhanced effects on $K$- and
$B$-meson FCNC observables vanish.  }

\vspace*{\fill}
\newpage

\setcounter{equation}{0}
\section{Introduction}
\label{sec:intro}

The  appearance of  too  large flavour-changing neutral-current (FCNC)
interactions of Higgs bosons to fermionic  matter is a generic feature
of SU(2)$_L\times $U(1)$_Y$  theories  with    two and more      Higgs
doublets.  Unless there is  a symmetry to  forbid these Higgs-mediated
FCNC    interactions  to occur    in   the  bare   Lagrangian of   the
model~\cite{GW}, their unsuppressed existence  will inevitably lead to
predictions for rare processes in the kaon  and $B$-meson systems that
violate     experimental   limits        by     several   orders    of
magnitude~\cite{GW,CS}.   In  the minimal realization of softly-broken
supersymmetry   (SUSY),   the   Minimal    Supersymmetric     Standard
Model~(MSSM),  the holomorphicity  of  the superpotential prevents the
occurrence of  Higgs-boson   FCNCs by coupling the  one  Higgs-doublet
superfield, $\widehat{H}_1$, to the down-quark  sector, and the  other
one, $\widehat{H}_2$,  to  the  up-quark sector.   However,  the above
holomorphic  property of  the   superpotential is  violated  by finite
radiative   (threshold)    corrections  due   to   soft  SUSY-breaking
interactions~\cite{Banks,RH}.   As a consequence, Higgs-mediated FCNCs
reappear at the one-loop  level, but are  naturally suppressed for low
and intermediate values of  $\tan\beta = \langle \widehat{H}_2 \rangle
/     \langle  \widehat{H}_1    \rangle$,     i.e.\  for    $\tan\beta
\stackrel{<}{{}_\sim} 20$.   For  larger values of $\tan\beta$,  e.g.\
$\tan\beta \stackrel{>}{{}_\sim} 30$, the FCNCs partially overcome the
loop  suppression factor $1/(16\pi^2)$  and  become phenomenologically
relevant~\cite{FCNC,BK},  especially   for   the  $K$-  and  $B$-meson
systems.

Recently, the   topic of  Higgs-boson  FCNCs in  the large-$\tan\beta$
limit    of       the    MSSM  has       received     much   attention
\cite{FCNC,BK,CGNW,FCNCCP,Frank,Isidori,Dedes1,ADKT,Ambrosio,Buras,MTW,IN}.
Several approaches have been devised  to implement the non-holomorphic
finite radiative corrections  into  the phenomenological  analysis  of
FCNC processes, such  as $K^0\bar{K}^0$ and $B^0\bar{B}^0$ mixings, $B
\to X_s \gamma$ and  $B_s\to  \ell^+\ell^-$.  In most cases,  however,
the   suggested approaches  to   threshold radiative  effects  involve
certain explicit or  implicit assumptions pertinent to the squark-mass
and the quark-Yukawa  structures of the  theory, such as the dominance
of  the top  quark in  the FCNC transition   amplitudes.  We  term the
latter assumption the   $t$-quark dominance hypothesis.  On  the other
hand,   some   of the approaches  neglect    higher-order terms in the
resummation of threshold  corrections  to $d$-quark Yukawa  couplings,
which become important in the large-$\tan\beta$ regime of the theory.

In this  paper we derive the effective  Lagrangian that properly takes
into   account the  resummation of higher-order   threshold effects on
Higgs-boson FCNC interactions to  down quarks.  To accomplish this  in
Section 2,  we avoid the imposition  of  particular assumptions on the
structure of the soft squark masses  and the quark-Yukawa couplings of
the   theory.   Moreover,  we  do  not  rely    on specific  kinematic
approximations   to  the    transition    amplitudes, such     as  the
aforementioned $t$-quark   dominance hypothesis  in  the  FCNC  matrix
elements.  In  our derivation  of  the effective  Lagrangian,  we also
consider the possibility of CP  violation through two sources: (i)~the
Cabibbo-Kobayashi-Maskawa (CKM) mixing matrix~\cite{CKM} and  (ii)~the
soft SUSY-violating    mass terms.  As   we explicitly  demonstrate in
Section~3, our resummed FCNC  effective Lagrangian  gives rise to  new
testable predictions  for   CP-conserving  as well    as  CP-violating
observables related  to the $K$- and  $B$-meson systems.   In the same
section,  we   qualitatively      discuss    the  implications      of
$\tan\beta$-enhanced    Higgs-mediated interactions    for the  direct
CP-violation  parameter $\epsilon'/\epsilon$   in  the  kaon   system.
Section~4 is devoted to our numerical analysis of a number of $K$- and
$B$-meson observables,  such  as $\Delta M_K$,   $\epsilon_K$, $\Delta
M_{B_{d,s}}$,   ${\cal  B}(  B_{d,s}  \to  \ell^+\ell^-   )$ and their
associated   leptonic  CP asymmetries.   In   particular, based on our
resummed   FCNC effective   Lagrangian,   we  are   able   to identify
configurations in the soft SUSY-breaking parameter  space, for which a
kind      of  a  Glashow--Iliopoulos--Maiani-cancellation    mechanism
(GIM)~\cite{GIM} becomes operative in the Higgs--$d$-quark sector.  As
a result, all Higgs-mediated, $\tan\beta$-enhanced effects on $K$- and
$B$-meson FCNC observables  vanish.  Finally, Section~5 summarizes our
conclusions.

\setcounter{equation}{0}
\section{Resummed FCNC effective Lagrangian}
\label{sec:sec2}

In  this  section,  we  derive  the general  form  for  the  effective
Lagrangian  of Higgs-mediated  FCNC interactions  in  the CP-violating
MSSM.    For   this   purpose,   we  also   consistently   resum   the
$\tan\beta$-enhanced  radiative   effects  on  the   $d$-quark  Yukawa
couplings~\cite{CGNW}.  First, we  analyze a simple soft SUSY-breaking
model    based    on     the    assumption    of    minimal    flavour
violation~\cite{BK,Isidori,Ambrosio,Buras},  where the  CKM  matrix is
the only source of flavour and CP violations. We find that even within
this minimal  framework, the usually  neglected $c$-quark contribution
to  Higgs-mediated  FCNC  interactions   may  be  competitive  to  the
$t$-quark one in certain regions  of the parameter space. After having
gained some insight from the  above considerations, we then extend our
resummed  effective Lagrangian  approach  to more  general cases  that
include a  non-universal or hierarchical  squark sector as well  as CP
violation originating  from the CKM matrix and  the soft SUSY-breaking
parameters.

\begin{figure}[t]

\begin{center}
\begin{picture}(320,150)(0,0)
\SetWidth{0.8}
 
\ArrowLine(0,70)(30,70)\ArrowLine(30,70)(60,70)
\ArrowLine(60,70)(90,70)\ArrowLine(90,70)(120,70)
\DashArrowArc(60,70)(30,90,180){3}\DashArrowArc(60,70)(30,0,90){3}
\DashArrowLine(60,130)(60,100){3}
\Text(60,70)[]{\boldmath $\times$}
\Text(0,65)[lt]{$d_L$}\Text(120,65)[rt]{$d_R$}
\Text(60,65)[t]{$\tilde{g}$}\Text(65,125)[l]{$\Phi^{0*}_{2}$}
\Text(37,100)[r]{$\tilde{d}^*_L$}
\Text(83,100)[l]{$\tilde{d}^*_R$}

\Text(60,25)[]{\bf (a)}

\ArrowLine(200,70)(230,70)\ArrowLine(230,70)(260,70)
\ArrowLine(260,70)(290,70)\ArrowLine(290,70)(320,70)
\DashArrowArc(260,70)(30,90,180){3}\DashArrowArc(260,70)(30,0,90){3}
\DashArrowLine(260,130)(260,100){3}
\Text(260,70)[]{\boldmath $\times$}
\Text(200,65)[lt]{$d_L$}\Text(320,65)[rt]{$d_R$}
\Text(245,65)[t]{$\tilde{h}^-_{2}$}
\Text(275,65)[t]{$\tilde{h}^-_{1}$}
\Text(265,125)[l]{$\Phi^{0*}_{2}$}
\Text(237,100)[r]{$\tilde{u}^*_R$}
\Text(283,100)[l]{$\tilde{u}^*_L$}

\Text(260,25)[]{\bf (b)}

\end{picture}
\end{center}
\vspace{-1.cm}
\noindent
\caption{\em Non-holomorphic radiative effects on the $d$-quark Yukawa
couplings induced by (a)~gluinos $\tilde{g}$ and (b)~charged Higgsinos
$\tilde{h}^-_{1,2}$.}\label{feyn}
\end{figure}
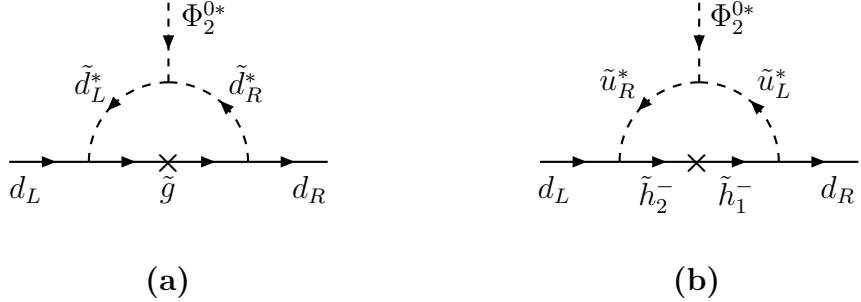

Before discussing the  most general  case,  let us  first consider the
following simple  form  for the effective Yukawa  Lagrangian governing
the    Higgs-mediated    FCNC     interactions    in    the      quark
sector~\cite{FCNC,BK}:
\begin{eqnarray}
  \label{lag}
-{\cal L}_Y\ =\ \bar{d}_R^0\, \hd \Big[\, \Phi^{0*}_1\: +\:
\Phi^{0*}_2\, \Big( \Eg\: +\: \Eu\, {\bf h_u^\dagger} 
\hu \Big)\,\Big]\, d_L^0\ +\
\Phi_2^0\, \bar{u}_R^0\, \hu\, u_L^0\ +\ {\rm H.c.} \,,
\end{eqnarray}
where $\Phi^0_{1,2}$ are the electrically neutral dynamical degrees of
freedom of the  two Higgs  doublets\footnote{Throughout the paper,  we
follow the notation and the CP-phase conventions of~\cite{CEPW}.}  and
the superscript `0' on the $d$-  and $u$-type quarks denotes fields in
the interaction basis.   In~(\ref{lag}), $\hd$ and $\hu$ are  $3\times
3$-dimensional      down-    and     up-quark   Yukawa       matrices,
and~\cite{RH,FCNC,BK}
\begin{eqnarray}
  \label{Eg}
\Eg &=& \asos\, 
\frac{2 \alpha_s}{3\pi}\, m_{\tilde{g}}^* \mu^*\, I(m_{\tilde{d}_L}^2,
m_{\tilde{d}_R}^2,|m_{\tilde{g}}|^2) \;, \\
  \label{Eu}
\Eu &=& \asos\, \frac{1}{16\pi^2}\,\mu^* A_U^*\, I(m_{\tilde{u}_L}^2,
m_{\tilde{u}_R}^2,|\mu|^2) 
\end{eqnarray}
are  finite non-holomorphic radiative effects  induced by the diagrams
shown in Fig.~\ref{feyn}. In the  above, the loop integral  $I(x,y,z)$
is given by
\begin{equation}
  \label{Ixyz}
I(x,y,z)\ =\ \frac{ xy \ln (x/y)\: +\: y z \ln (y/z)\: +\: x z \ln (z/x)}
{(x-y)\,(y-z)\,(x-z)}\ ,
\end{equation}
with $I(x,x,x) = 1/(2x)$.  To keep  things simple in the beginning, we
assume  that     in~(\ref{Eg})  and~(\ref{Eu}),  the    bilinear  soft
SUSY-breaking  masses    of  the   squarks,   $m^2_{\tilde{u}_{L,R}}$,
$m^2_{\tilde{d}_{L,R}}$, and  the trilinear soft Yukawa couplings $A_U
=    A_{u,d}$ are   flavour-diagonal  and   universal    at the   soft
SUSY-breaking scale $M_{\rm  SUSY}$.  We  also neglect the  left-right
mixing    terms                  $\tilde{u}_L$-$\tilde{u}_R$       and
$\tilde{d}_L$-$\tilde{d}_R$   in  the  squark    mass matrices.    The
consequences of relaxing the above assumptions will be discussed later
on.

{}From~(\ref{lag}), we can easily  write down the effective Lagrangian
relevant to the effective $d$- and $u$-type quark masses:
\begin{eqnarray}
  \label{masslag}
-{\cal L}_{\rm mass}\ =\ \frac{v_1}{\sqrt{2}}\, \bar{d}_R^0\,  \hd \Big[\,
\asos\: +\: \tan\beta\, \Big(\,\Eg\: +\: \Eu\, 
{\bf h_u^\dagger} \hu\, \Big)\, \Big]\, d_L^0\ +\
\frac{v_2}{\sqrt{2}}\, \bar{u}_R^0\, \hu\, u_L^0\ +\ {\rm H.c.}
\end{eqnarray}
Our next step is to redefine the quark fields as follows:
\begin{eqnarray}
  \label{rotation}
u_L^0\ =\ \ULQ\, u_L \,,\quad  d_L^0\ =\ \ULQ\, \Vckm\, d_L\,,\quad 
u_R^0\ =\ \URu\, u_R\,,\quad d_R^0\ =\ \URd\, d_R \,,
\end{eqnarray}
where  $\ULQ$, $\URu$, $\URd$  and $\Vckm$ are 3-by-3 unitary matrices
that relate  the weak   to mass  eigenstates of  quarks.    Evidently,
$\Vckm$   is     by     construction the      physical   CKM   matrix.
Substituting~(\ref{rotation}) into (\ref{masslag}) yields
\begin{eqnarray}
  \label{diago}
-{\cal L}_{\rm mass}  \!\!\!&=&\!\!\! 
\frac{v_1}{\sqrt{2}}\, \bar{d}_R\,  \URd^\dagger \hd \ULQ\,
\Big[\, \asos\: +\: \tan\beta\, \Big(\,\Eg\: + \: \Eu
|\hhu|^2\,\Big)\, \Big]\, \Vckm\, d_L\
+\ \frac{v_2}{\sqrt{2}}\, \bar{u}_R\, \hhu\, u_L\ +\ {\rm H.c.}
\nonumber \\[3mm] 
&=& \bar{d}_R\, \Md\, d_L\ +\ \bar{u}_R\, \Mu\, u_L\ +\ {\rm H.c.}\;,
\end{eqnarray}
where   $\Md$  and $\Mu$  are the    physical  $d$- and $u$-quark mass
matrices, respectively. Consistency of~(\ref{diago}) implies
\begin{eqnarray}
  \label{Mu}
\Mu &=& \frac{v_2}{\sqrt{2}}\, \hhu \;,\\
  \label{hd}
\URd^\dagger \hd \ULQ &=& \frac{\sqrt{2}}{v_1}\, \Md \Vckm^\dagger\,
\Rh^{-1}\;, 
\end{eqnarray}
with 
\begin{eqnarray}
  \label{Rhat}
\Rh \ =\  \asos\: +\: \Eg \tan\beta \: +\: \Eu \tan\beta\, |\hhu|^2  \;.
\end{eqnarray}
Notice    that  (\ref{hd})    plays  the  r\^ole     of a  re-defining
(renormalization) condition for the $d$-quark Yukawa couplings, in the
process of resumming higher-order radiative corrections.  Observe also
that the matrix $\Rh$ cannot be zero, as this would result in massless
$d$ quarks.

With the help  of~(\ref{rotation}) and~(\ref{hd}), we can now  express
our original  Yukawa  Lagrangian  (\ref{lag}) in   terms  of the  mass
eigenstates $d_{L,R}$ and $u_{L,R}$ in a resummed form:
\begin{eqnarray}
  \label{LYeff}
-{\cal L}_Y &=& \frac{\sqrt{2}}{v_1}\,  
\bar{d}_R\,  \Md \Vckm^\dagger \Rh^{-1} 
\Big[\, \Phi_1^{0*}\: +\: \Phi_2^{0*} \Big(\Eg\: +\: \Eu |\hhu|^2\Big)\, \Big]
\Vckm\, d_L\ +\ \Phi_2^0\, \bar{u}_R \hhu u_L\ +\ {\rm H.c.} \nonumber\\
&=& \frac{\sqrt{2}}{v_2}\,
\Big( \tan\beta\,\Phi_1^{0*}\: -\: \Phi_2^{0*} \Big)\,
\bar{d}_R\,  \Md \Vckm^\dagger \Rh^{-1}\Vckm\, d_L\ +\
\frac{\sqrt{2}}{v_2}\, \Phi_2^{0*}\, \bar{d}_R\,  \Md\, d_L\nonumber\\
&&+\ \Phi_2^0\, \bar{u}_R \hhu u_L\ +\ {\rm H.c.}
\end{eqnarray}
In deriving  the last equality  in (\ref{LYeff}), we have employed the
relation: $\Rh^{-1}     (\Eg   +  \Eu   |\hhu|^2  )\  =\     (\asos  -
\Rh^{-1})/\tan\beta$.
 
It is very illuminating to see how the FCNC part of (\ref{LYeff})
compares  with   the    literature,  e.g.\ with    that   obtained  in
Ref.~\cite{BK}.  To  this end,  let  us first  assume that $\asos +\Eg
\tan\beta \ne 0$ and decompose $\Rh^{-1}$ as follows:
\begin{eqnarray}
  \label{Rhdec}
\Rh^{-1}\ =\ \frac{\asos}{\asos+\Eg \tan\beta}\ +\ \chi_{\rm FC}\; ,  
\end{eqnarray}
where  $\chi_{\rm  FC}$ is  the  diagonal  matrix
\begin{equation}
  \label{chiFC}
\chi_{\rm FC}\ =\ -\,
\frac{\Eu\, |{\bf \hat{h}_u}|^2\, \tan\beta}
{(\asos\, +\, \Eg \tan\beta)\, \Rh}\ \; .  
\end{equation}
Making use  of the above  linear decomposition  of $\Rh^{-1}$  and the
unitarity  of the CKM  matrix in~(\ref{LYeff}),  the  FCNC part of our
resummed effective Lagrangian reads
\begin{eqnarray}
  \label{cbk}
-{\cal L}_{\rm FCNC} \!\!&=&\!\! 
 \frac{\sqrt{2}}{v_2}\, \Big( \tan\beta\,\Phi_1^{0*}\: -\: 
\Phi_2^{0*} \Big)\, \bar{d}_{iR}\, m_{d_i}\,
\bigg(\, V_{ti}^* \chi_{\rm FC}^{(t)} V_{tj}\: +\: V_{ci}^* \chi_{\rm FC}^{(c)}
V_{cj}\: +\: V_{ui}^* \chi_{\rm FC}^{(u)} V_{uj}\, \bigg)  d_{jL}
\,,\nonumber\\ &&
\end{eqnarray}
where $\chi_{\rm FC}^{(u,c,t)}$ are the diagonal entries of $\chi_{\rm
FC}$ and   summation  over $i,j =  d,s,b$   is  understood.   The term
proportional   to $\chi_{\rm    FC}^{(t)}$    gives  the  top    quark
contribution,  which  is  the   result   of~\cite{BK}  and  subsequent
articles~\cite{Isidori,Ambrosio,Buras}.   However,  we   should remark
here that the frequently-used top-quark dominance approximation cannot
be    justified   from    considerations   based   only   on   minimal
flavour-violation   models~\cite{BK,Isidori,Ambrosio,Buras}.  In fact,
the other terms in~(\ref{cbk})  and especially the one proportional to
$\chi_{\rm  FC}^{(c)}$  due  to the   charm-quark  contribution become
rather important in the  limit $\asos + \Eg  \tan\beta \to 0$. In this
limit, the  singularity  in $\chi_{\rm FC}^{(t)}$ is  canceled against
the  singularities of $\chi_{\rm  FC}^{(c)}$ and $\chi_{\rm FC}^{(u)}$
as a result of the  unitarity of $\Vckm$.   In this context, we should
note  that the limit $\asos +  \Eg \tan\beta \to  0$ is not attainable
before the theory itself reaches a non-perturbative regime.  Requiring
that all $d$-quark Yukawa couplings are  perturbative, we can estimate
the lower   bound,  $|\asos  +  \Eg   \tan\beta| \stackrel{>}{{}_\sim}
2.5\times  10^{-2}$, for $\tan\beta=50$.\footnote{To obtain this lower
limit, we simply take the trace of the square of~(\ref{hd}) and demand
that ${\rm   Tr}|\hhd|^2    <  3$, or    equivalently   ${\rm  Tr}   [
(\Rh^{-1})^\dagger V |\Md|^2 V^\dagger \Rh^{-1} ] <  3 v_1^2/ 2$.  The
latter    implies    that       $|\asos\:    +\:    \Eg    \tan\beta|\
\stackrel{>}{{}_\sim}\ (m_s  \tan\beta)/(\sqrt{3}\,  m_t)\  =\ 5\times
10^{-4} \tan\beta$.   Finally, it is amusing to  notice that if $\Im m
\Eg  =0$   and  $\Re e  \Eg <    0$,  a perturbative  upper  bound  on
$\tan\beta$, $\tan\beta \stackrel{<}{{}_\sim} 1/|\Eg|$, may be derived
beyond the  tree level.}  Although  $|\asos\: +\: \Eg \tan\beta|$ must
not vanish  in  perturbation theory, it can  be  sufficiently close to
zero, so that the  $c$-quark contribution becomes competitive with the
$t$-quark one.
 
So far, we  have  assumed that the radiatively-induced  matrices $\Eg$
and $\Eu$   in    the  effective Yukawa   Lagrangian~(\ref{lag})   are
proportional to the unity matrix.  However, this assumption of flavour
universality is rather specific. It  gets generally invalidated by the
mixing of the squark generations, the  soft trilinear Yukawa couplings
and  renormalization-group (RG)  running  of   the  soft SUSY-breaking
parameters   from the unification to   the  low-energy scale.  In this
respect, the  minimal-flavour-violation   hypothesis, although  better
motivated, should  also be viewed  as a  particular way from minimally
departing from universality.

Nevertheless, given that threshold radiative effects  on the up sector
are  negligible, especially  for large  values  of  $\tan\beta$, it is
straightforward  to  derive   the   general resummed   form   for  the
Higgs-mediated  FCNC  effective  Lagrangian.   Starting  from  general
non-diagonal  matrices $\Egt$ and  $\Eut$ in~(\ref{lag}) and following
steps very  analogous to  those from~(\ref{lag})  to~(\ref{LYeff}), we
arrive  at  the same form as   in~(\ref{LYeff}) for  the resummed FCNC
effective Lagrangian, but with $\Eg$, $\Eu$ and $\Rh$ replaced by
\begin{eqnarray}
  \label{Epars} 
\Egb &=& \ULQ^\dagger \Egt \ULQ\,, \qquad\qquad 
\Eub\ =\ \ULQ^\dagger \Eut \ULQ\,, \\[2mm] 
  \label{Ro}
\Ro &=& \asos\: +\: \tan\beta\, \Big(\Egb \:  +\: \Eub\, |\hhu|^2\:
+\: \dots \Big) \,,
\end{eqnarray}
where the  ellipses   in~(\ref{Ro})  denote  additional   (generically
sub-dominant)   threshold  effects~\cite{comment}.   Notice  that  the
unitary matrix $\ULQ$ in~(\ref{Epars}), which  is only constrained  by
the  relation $|\hhu|^2=  \ULQ^\dagger  {\bf  h_u^\dagger}  \hu \ULQ$,
introduces additional  non-trivial  model  dependence in the  matrices
$\Egb$,   $\Eub$ and $\Ro$.  In other   words, the  presence of $\ULQ$
reflects  the fact  that   the $3\times 3$-dimensional matrices  ${\bf
h_u^\dagger}  \hu$, $m^2_{\tilde{u}_L}$ and $m^2_{\tilde{d}_L}$ cannot
be diagonalized simultaneously,  without generating FCNC  couplings in
other  interactions   in   the  MSSM    Lagrangian,   e.g.\ in     the
$\tilde{W}_3$-$u_L$-$\tilde{u}_L$                                  and
$\tilde{W}^-$-$u_L$-$\tilde{d}_L$      couplings.   Moreover, even  in
minimal    flavour-violating scenarios,   the $3\times  3$-dimensional
matrix $\Ro$    may  generally contain    additional radiative effects
proportional to $\ULQ^\dagger  {\bf h_d^\dagger} \hd \ULQ$ induced  by
RG running  of the squark  masses  from  the unification  to the  soft
SUSY-breaking scale   $M_{\rm  SUSY}$.   These contributions   can  be
resummed individually by  taking appropriately the hermitian square of
the   modified~(\ref{hd})    and    solving  for   $\ULQ^\dagger  {\bf
h_d^\dagger}   \hd  \ULQ$.   This last  step  may  involve  the use of
iterative or other numerical methods.

In the  general case  of a  non-universal  squark sector, the resummed
FCNC couplings of  the Higgs bosons  to down-type quarks can always be
parameterized   in terms of  a  well-defined set of  parameters at the
electroweak scale. In the  weak basis, in which  $\ULQ = {\bf 1}$, the
set of input parameters consists of: (i) the soft squark mass matrices
$m^2_{\tilde{u}_{L,R}}$,        $m^2_{\tilde{d}_{L,R}}$    and    soft
Yukawa-coupling   matrices $A_{u,d}$; (ii)    the $u$-  and  $d$-quark
masses; (iii) the CKM mixing matrix~$\Vckm$.

In  our last step in  deriving the resummed FCNC effective Lagrangian,
we  express  the Higgs fields $\Phi^0_{1,2}$  in   terms of their mass
eigenstates $H_{1,2,3}$ and the neutral would-be Goldstone boson $G^0$
in  the   presence  of  CP   violation~\cite{APLB}.    Following   the
conventions  of~\cite{CEPW}, we  relate the weak   to mass eigenstates
through the linear transformations:
\begin{eqnarray} 
  \label{CPtr}
\Phi_1^0 &=& \frac{1}{\sqrt{2}}\, \Big[\, O_{1i}\, H_i\ +\ i\, 
\Big( \cos\beta ~G^0\: -\: \sin\beta ~O_{3i}\, H_i\,\Big)\, \Big]\,,
  \nonumber \\[3mm]
\Phi_2^0 &=& \frac{1}{\sqrt{2}}\, \Big[\, O_{2i}\, H_i\ +\ i\,
\Big( \sin\beta ~G^0\: +\: \cos\beta ~O_{3i}\, H_i\,\Big)\, \Big] \,,
\end{eqnarray}
where  $O_{ij}$  is a 3-by-3   orthogonal   matrix that accounts   for
CP-violating Higgs-mixing   effects~\cite{PW}.  If we   substitute the
weak  Higgs  fields   $\Phi^0_{1,2}$  by virtue  of~(\ref{CPtr})  into
(\ref{LYeff}), we obtain the general resummed effective Lagrangian for
the diagonal as well as off-diagonal Higgs interactions to $d$ quarks,
\begin{eqnarray}
  \label{master}
{\cal L}_{H_i\bar{d}   d'}\  =\ -\,  \frac{g_w}{2  M_W}\, \sum_{i=1}^3\,
H_i\:  \bar{d}\, \Big(\, \Md\, {{\bf g}}_{H_i\bar{d}  d' }^L\, P_L\ +\
{{\bf g}}_{H_i\bar{d}  d' }^R\, {\bf \hat{M}_{d'}} P_R\, \Big)\, d' \,,
\end{eqnarray}
where $P_{L\, (R)} = [1 -(+)\, \gamma_5 ]/2$ and
\begin{eqnarray}
  \label{couplings}
{{\bf g}}_{H_i\bar{d} d' }^L \!\!&=&\!\! 
\Vckm^\dagger\Ro^{-1}\Vckm\, \frac{O_{1i}}{\cos\beta}\ +\
\bigg(\asos\: -\: \Vckm^\dagger\Ro^{-1}\Vckm\,\bigg)\,\frac{O_{2i}}{\sin\beta}
\ -\ i\, \bigg( \asos\: -\: \frac{1}{\cos^2 \beta}\, 
\Vckm^\dagger\Ro^{-1}\Vckm\, \bigg)\, \frac{O_{3i}}{\tan\beta}\ ,
\nonumber\\[3mm]
{{\bf g}}_{H_i\bar{d} d' }^R \!\!&=&\!\!  \big(\,{{\bf g}}_{H_i\bar{d} d'
  }^L\,\big)^\dagger\, .
\end{eqnarray}
The $3\times 3$-dimensional  matrix $\Ro$ in~(\ref{couplings}),  which
resums all  $\tan\beta$-enhanced  finite radiative  effects, is  given
by~(\ref{Ro}).  Equation~(\ref{master}), along with~(\ref{couplings}),
constitutes the major  result  of  the  present paper,  which will  be
extensively used in our phenomenological discussions in Section~3.

Finally,  let us   summarize the  most   important  properties  of the
resummed effective Lagrangian~(\ref{master}):

\begin{itemize}

\item[(i)] The  FCNC  interactions in~(\ref{master}) are  described by
$\tan^2\beta$-enhanced  terms that are  proportional  to $O_{1i}$  and
$O_{3i}$ and  to  $\Vckm^\dagger \Ro^{-1} \Vckm$ in~(\ref{couplings}).
These  $\tan^2\beta$-enhanced  FCNC terms properly   take into account
resummation,\footnote{In addition to the non-holomorphic contributions
we  have  been considering  here,   there  are in  general holomorphic
radiative effects on the $\Phi^0_1$ coupling to  $d$ quarks which have
an    analogous matrix structure,   i.e.\ ${\bf  \varepsilon_g} + {\bf
\varepsilon_u}   |\hhu|^2$.   These additional   holomorphic terms are
generally small,  typically of   order $10^{-2}$,  and  only  slightly
modify the form of the matrix ${\bf  R}$ to: ${\bf R}  = {\bf 1}\, +\,
({\bf 1} + {\bf \varepsilon_g} + {\bf  \varepsilon_u} |\hhu |^2 )^{-1}
(\Egb\, +\,  \Eub   |\hhu   |^2)\,  \tan\beta$.   Obviously,  such   a
modification  does  not  alter   the  general form   of the  effective
couplings in~(\ref{couplings}) and is beyond the one-loop order of our
resummation.  Therefore, these  additional small holomorphic terms can
be safely  neglected.}    non-universality in the  squark  sector  and
CP-violating effects.

\item[(ii)] The resummation matrix  ${\bf R}$ controls the strength of
the  Higgs-mediated FCNC  effects.   For instance,  if  ${\bf   R}$ is
proportional to    unity,   then   a   kind  of   a   GIM-cancellation
mechanism~\cite{GIM}      becomes   operative    and  the  Higgs-boson
contributions to all FCNC observables vanish identically in this case.
Furthermore, as  well as the top quark,  the other two lighter up-type
quarks  can  give   significant  contributions to FCNC  transition
amplitudes, which are naturally included in~(\ref{master}) through the
resummation matrix ${\bf R}$.

\item[(iii)] In the CP-invariant   limit of the theory,  the effective
couplings ${{\bf  g}}_{H_i\bar{d} d' }^{L,R}$ are  either pure real or
pure imaginary numbers. Moreover, in   the limit $\Vckm \to {\bf  1}$,
the    effective    Lagrangian   (\ref{master})     of the    diagonal
Higgs-couplings to down quarks is in  excellent agreement with the one
presented in~\cite{CEPW,APEDM}. 

\item[(iv)] If $M_{H^+} \sim   M_{H_{2,3}} \gg M_{H_1}$, one can  show
that  $O_{11}  \approx  \cos\beta$,   $O_{21} \approx   \sin\beta$ and
$O_{31}\approx  0$.  In this case,   the $H_1$-coupling to $d$  quarks
becomes SM like  and so $H_1$-mediated  FCNC effects  are getting 
suppressed.  Instead, the FCNC couplings of the heavy $H_{2,3}$ bosons
to $d$-type quarks retain their  $\tan\beta$-enhanced strength in  the
above kinematic region.

\item[(v)]  The  one-loop resummed effective Lagrangian~(\ref{master})
captures the major  bulk  of the one-loop radiative  effects~\cite{LN}
for  large      values   of  $\tan\beta$,   e.g.\     for   $\tan\beta
\stackrel{>}{{}_\sim} 40$, and for a  soft SUSY-breaking scale $M_{\rm
SUSY}$  much  higher  than   the   electroweak scale~\cite{CGNW}.   In
addition, (\ref{master})  is  only valid in   the limit  in which  the
four-momentum of the $d$ quarks and Higgs bosons  in the external legs
is  much smaller  than    $M_{\rm  SUSY}$.  This   last   condition is
automatically   satisfied  in our   computations   of low-energy  FCNC
observables.

\end{itemize}

In the next section,  we  will study  in detail  the  phenomenological
consequences of the  $\tan\beta$-enhanced   FCNC effects mediated   by
Higgs bosons on rare processes and CP  asymmetries related to the $K$-
and $B$-meson systems.

\setcounter{equation}{0}
\section{Applications to $K$- and $B$-meson systems}

We   shall    now  analyze  the   impact    of our  resummed effective
Lagrangian~(\ref{master})   for  Higgs-mediated  FCNC  interactions on
representative   $K$- and  $B$-meson   observables.  For comprehensive
reviews   on  $K$- and   $B$-meson   physics,  we  refer    the reader
to~\cite{kaonrev,Neubert,Branco}.

\subsection{$\DmK$, $\ek$ and $\epsilon'/\epsilon$}

Our starting point is the effective Hamiltonian for the $\Delta S = 2$
interactions,
\begin{eqnarray}
  \label{DS2}
H_{\rm eff}^{\Delta S=2}\ =\ 
\frac{G_F^2}{16\pi^2}\, M_W^2\, \sum_i\, C_i(\mu)\, Q_i (\mu ) \;,
\end{eqnarray}
where $G_F=1.16639\times    10^{-5}~{\rm  GeV}^{-2}$ is     the  Fermi
constant, and $C_i   (\mu)$ are  the  scale-dependent Wilson  coefficients
associated to the $\Delta S=2$  quark-dependent operators $Q_i$.  Note
that the CKM  matrix elements in~(\ref{DS2})  have  been absorbed into
the  Wilson coefficients.  The $\Delta  S=2$  operators  $Q_i$ may  be
summarized as follows:
\begin{eqnarray}
  \label{operbasis}
Q_1^{\rm VLL} &=&(\bar{s}\gamma_\mu P_L d)(\bar{s}\gamma^\mu P_L d)\,,
\qquad\!
Q_1^{\rm VRR}\ =\ (\bar{s}\gamma_\mu P_R d)(\bar{s}\gamma^\mu P_R d)\,,
\nonumber\\[3mm]
Q_1^{\rm LR} &=&(\bar{s}\gamma_\mu P_L d)(\bar{s}\gamma^\mu P_R d)\, ,
\qquad\
Q_2^{\rm LR}\ =\ (\bar{s} P_L d)(\bar{s} P_R d)\,,
\nonumber \\[3mm]
Q_1^{\rm SLL} &=&(\bar{s} P_L d)(\bar{s} P_L d)\, ,\qquad\qquad
Q_1^{\rm SRR}\ =\ (\bar{s} P_R d)(\bar{s} P_R d)\,,
\nonumber \\[3mm]
Q_2^{\rm SLL} &=&(\bar{s}\sigma_{\mu\nu} P_L d)(\bar{s}\sigma^{\mu\nu} 
P_L d)\, ,\quad
Q_2^{\rm SRR}\ =\ (\bar{s}\sigma_{\mu\nu} P_R d)(\bar{s}\sigma^{\mu\nu} 
P_R d) \;,
\end{eqnarray}
with  $\sigma_{\mu\nu}=\frac{1}{2}[\gamma_\mu,\gamma_\nu]$. Here, much
of our discussion and notation follows Ref.~\cite{Jager}.  

It now proves convenient to  decompose both the $K^0$-$\bar{K}^0$ mass
difference $\DmK$ and  the known  CP-violating mixing parameter  $\ek$
into a SM and a SUSY contribution:
\begin{eqnarray}
\Delta M_K &=& M_{K_L} \: -\: M_{K_S}\ =\ 
\Delta M_K^{\rm SM} + \Delta M_K^{\rm SUSY} \;,\nonumber\\ 
\epsilon_K &=& \epsilon_K^{\rm SM} + \epsilon_K^{\rm SUSY} \;.
\end{eqnarray}
To a good approximation, one has
\begin{eqnarray}
\Delta M_K^{\rm SM,\,SUSY} &=& 2\, \Re e\, 
\bra{\bar{K}^0}\,
H_{\rm eff}^{\Delta S=2}\, \ket{K^0}_{\rm SM,\,SUSY} \;, \\[3mm]
\epsilon_K^{\rm SM,\,SUSY} &=& \frac{{\rm exp}(i\pi/ 4)}{\sqrt{2}\, 
\Delta M_K}\ \Im m\,  \bra{\bar{K}^0}\, H_{\rm eff}^{\Delta S=2}\,
\ket{K^0}_{\rm SM,\, SUSY} \; .
\end{eqnarray}
The SUSY  contribution to  the matrix   element $\bra{\bar{K}^0}H_{\rm
eff}^{\Delta   S=2}   \ket{K^0}_{\rm   SUSY}$ may  be   written   down
as~\cite{Jager,Chankowski}
\begin{eqnarray}
  \label{Kme}
\bra{\bar{K}^0}\, H_{\rm eff}^{\Delta S=2}\, \ket{K^0}_{\rm SUSY} &=&
\frac{G_F^2}{12\pi^2}\, M_W^2 m_K F_K^2 \eta_2 \hat{B}_K\, \Big[\,
\bar{P}_1^{\rm VLL}\, \Big(\, C_1^{\rm VLL}\: +\:  C_1^{\rm VRR}
\,\Big)\nonumber \\[3mm]
&&+\: \bar{P}_1^{\rm LR}\, C_1^{\rm LR}\ 
+\ \bar{P}_2^{\rm LR}\, C_2^{\rm LR} \nonumber\\[3mm]
&&+\: \bar{P}_1^{\rm SLL}\, \Big(\, C_1^{\rm SLL}\: +\: C_1^{\rm SRR}\, \Big)\
+\ \bar{P}_2^{\rm SLL}\, 
\Big(\, C_2^{\rm SLL}\: +\: C_2^{\rm SRR}\, \Big)\, \Big]\,,\qquad 
\end{eqnarray}
where  $m_K=498$   MeV,   $F_K=160$~MeV  and the
  $\bar{P}$'s  are    the
next-to-leading order (NLO)  QCD  factors  that include  the  relevant
hadronic                                                        matrix
elements~\cite{Jager,Ciuchini,Misiak,Becirevic,Chankowski}.     At the
scale $\mu=2$ GeV, they are given by~\cite{Jager}
\begin{eqnarray}
  \label{Pbars}
\bar{P}_1^{\rm VLL} = 0.25\,,\quad
\bar{P}_1^{\rm LR} = -18.6\,, \quad
\bar{P}_2^{\rm LR} = 30.6\,, \quad
\bar{P}_1^{\rm SLL} = -9.3\,, \quad 
\bar{P}_2^{\rm SLL} = -16.6\;.
\end{eqnarray}
On   obtaining~(\ref{Pbars}),  we  have  used  the   numerical values:
$\eta_2=0.57$ and $\hat{B}_K=0.85 \pm 0.15$.

{}From studies in  the     CP-conserving MSSM with  minimal    flavour
violation~\cite{Chankowski,Isidori,Buras},  it is known that for large
values   of   $\tan\beta   \stackrel{>}{{}_\sim}   40$,   the dominant
contribution     to    $\bra{\bar{K}^0}H_{\rm    eff}^{\Delta     S=2}
\ket{K^0}_{\rm SUSY}$  comes  from  Higgs-mediated two-loop double
penguin (DP)   diagrams proportional to   $C_2^{\rm  LR}$.  Within the
framework     of   our    large-$\tan\beta$-resummed  FCNC   effective
Lagrangian~(\ref{master}) that   includes  CP  violation,  the  Wilson
coefficients due to DP graphs are found to be
\begin{eqnarray}
  \label{dpkaon}
C_1^{\rm SLL\,(DP)} &=& -\,\frac{16\pi^2 m_s^2}{\sqrt{2}\, G_F M_W^2}\,
\sum_{i=1}^3\, \frac{g_{H_i\bar{s}d}^L\, g_{H_i\bar{s}d}^L }{M_{H_i}^2} \;, 
\nonumber \\[3mm]
C_1^{\rm SRR\,(DP)} &=& -\,\frac{16\pi^2 m_d^2}{\sqrt{2}\, G_F M_W^2}\,
\sum_{i=1}^3\, \frac{g_{H_i\bar{s}d}^R\, g_{H_i\bar{s}d}^R }{M_{H_i}^2} \;,  
\nonumber \\[3mm]
C_2^{\rm LR\,(DP)} &=& -\,\frac{32\pi^2 m_s m_d}{\sqrt{2}\, G_F M_W^2}\,
\sum_{i=1}^3\, \frac{g_{H_i\bar{s}d}^L\, g_{H_i\bar{s}d}^R }{M_{H_i}^2} \;,
\end{eqnarray}
where the $\tan^2\beta$-enhanced couplings $g_{H_i\bar{s}d}^{L,R}$ may
be  evaluated   from~(\ref{couplings}).    Note  that  the DP   Wilson
coefficients in~(\ref{dpkaon}) exhibit a $\tan^4\beta$-dependence and,
although being two-loop  suppressed, they become  very significant for
large values of $\tan\beta \stackrel{>}{{}_\sim} 40$.

In addition to the  aforementioned DP contributions due to Higgs-boson
exchange   graphs,  there  exist  relevant  one-loop  contributions to
$\bra{\bar{K}^0}H_{\rm eff}^{\Delta S=2}\ket{K^0}_{\rm SUSY}$ at large
$\tan\beta$: (i) the $t$-$H^\pm$ box contribution to $C_2^{\rm LR}$ of
the  two-Higgs-doublet-model   (2HDM)  type, and  (ii)  the   one-loop
chargino-stop box diagram contributing to  $C_1^{\rm SLL}$.  The first
contribution~(i)  becomes significant,  up   to  10\%,  only  in   the
kinematic region $M_{H^\pm}\approx   m_t$.  In this  case,  to a  good
approximation,     $C_1^{\rm          SLL}$         may    be    given
by~\cite{Chankowski,Isidori}
\begin{eqnarray}
  \label{2HDM}
C_2^{\rm LR\,(2HDM)} \approx -\,\frac{2 m_s m_d}{M_W^2}\, 
(V_{ts}^*V_{td})^2\, \tan^2\beta \;. 
\end{eqnarray}
Note that   the light-quark   masses contained  in~(\ref{dpkaon})  and
(\ref{2HDM})  are running  and are  evaluated   at the  top-quark mass
scale, i.e.\ $m_s(m_t)  \simeq 61$ MeV, $m_d (m_t)  \simeq 4$ MeV. The
second contribution~(ii)  becomes non-negligible only for small values
of  the $\mu$-parameter~\cite{Isidori,Chankowski},  i.e.\ for  $|\mu |
\stackrel{<}{{}_\sim} 200$~GeV.

In  view of  the above  discussion,  the kinematic parameter range  of
interest to us is:
\begin{eqnarray}
  \label{par}
M_{\rm SUSY},\; \mu\ \gg\ m_t\, , \qquad \tan\beta\ 
\stackrel{>}{{}_\sim}\ 40\,, 
\end{eqnarray}
including   the  case   $M_{H^\pm}  \approx    m_t$,   for which   the
Higgs-mediated   DP  effects   can    dominate  the  $K^0$-$\bar{K}^0$
transition amplitude.  Thus, taking also into account the sub-dominant
2HDM contribution~(\ref{2HDM}), formula~(\ref{Kme}) simplifies to
\begin{eqnarray}
  \label{Ksusy}
\bra{\bar{K}^0}\,H_{\rm eff}^{\Delta S=2}\,\ket{K^0}_{\rm SUSY}
\!\!&=&\!\! 
4.6\times 10^{-11} {\rm ~GeV}\ 
\bigg(\, \frac{F_K}{160 {\rm ~MeV}}\, \bigg)^2\,
\bigg(\,\frac{\eta_2}{0.57}\, \bigg)\,
\bigg(\,\frac{\hat{B}_K}{0.85}\, \bigg)\nonumber\\[3mm]
&&\hspace{-2.5cm} \times\,
\biggl [\, 30.6\, \biggl (\, C_2^{\rm  LR\,(DP)}\: +\: 
C_2^{\rm  LR\,(2HDM)} \,\biggr )\ -\ 9.3\, 
\biggl (\, C_1^{\rm  SLL\,(DP)}\: +\: 
C_1^{\rm  SRR\,(DP)}\, \biggr)\, \biggr ] \,.\qquad 
\end{eqnarray}
Observe  that the Wilson coefficients $C_1$  and $C_2$ contribute with
opposite       signs     to   $\bra{\bar{K}^0}\,H_{\rm    eff}^{\Delta
S=2}\,\ket{K^0}_{\rm  SUSY}$. Based   on~(\ref{Ksusy}), we will   give
numerical estimates of $\Delta M_K$ and $\epsilon_K$ in Section 4.

We now turn   our   attention  to  the  computation  of   the   direct
CP-violation   parameter   $\epsilon'/\epsilon$ in the   kaon  system,
induced by CP-violating Higgs-mediated  FCNC interactions.  In the SM,
unlike the $Z$-penguin graphs, the Higgs-penguin contribution to $K\to
\pi\pi$, which is proportional to the operator
\begin{eqnarray}
 \label{qhiggs}
Q_H\ =\ (\bar{s} P_L d)\, \sum_{q=u,d,s}\, (\bar{q}q) \;,
\end{eqnarray}
has a  suppressed  Wilson coefficient proportional to  $m^2_q/M^2_H$,
where the  SM Higgs-boson mass $M_H$ is  subject into the experimental
bound: $M_H \stackrel{>}{{}_\sim} 114$~GeV.  One  might even think  of
the possibility that  the operator $Q_H$ in~(\ref{qhiggs}), which  has
enhanced Wilson  coefficients for $q=c,b$,  mixes with the gluonic and
electroweak  penguins,  as  well as  with  the  other  basis operators
in~(\ref{operbasis}).     However,  as    was  already pointed     out
in~\cite{Buchalla}, this is not the case,  and so the SM-Higgs penguin
effects remain negligible.

The  situation changes  drastically  in  the   MSSM with explicit   CP
violation, since the   Higgs-boson FCNC couplings to  down-type quarks
are substantially enhanced by $\tan^2\beta$-dependent terms, for large
values of  $\tan\beta$.    Furthermore, besides  the CKM  phase,   the
presence of complex  soft SUSY-breaking masses with large CP-violating
phases   may further   increase    the Higgs-boson   FCNC  effects  on
$\epsilon'/\epsilon$.  In fact, we note  that soft CP-odd phases could
even be   the   only   source~\cite{Banks}  to  account  for    direct
CP violation.

To reliably  estimate the new SUSY  effect on $\epsilon'/\epsilon$ due
to Higgs-boson  FCNC  interactions,   we  normalize  each   individual
contribution with respect to the dominant SM contribution arising from
the   operator $Q_6  = \sum_{q=u,d,s}   (\bar{s}   P_R q)(\bar{q}  P_L
d)$~\cite{EAP,Bosch}, with Wilson coefficient $y_6$, viz.\
\begin{eqnarray}
  \label{ee6}
\epe \ =\ \biggl ( \epe \biggr )_6\, \Big(\, 
\Omega_{\rm SM}\ +\ \Omega_{\rm SUSY}\ +\ 
\Omega_{\rm SUSY}^{\rm Higgs}\, \Big)\ .
\end{eqnarray}
In the MSSM with     minimal flavour violation, the  non-Higgs    SUSY
contribution  $\Omega_{\rm  SUSY}$ is   small~\cite{Guidice}.  Sizable
contributions  may  be obtained  if  one  relaxes the  assumptions  of
universality   and      CP    conservation   in         the     squark
sector~\cite{epeSUSY,Kagan:1999iq}.

Here, we compute  a novel contribution to $\epsilon'/\epsilon$, namely
the   quantity $\Omega_{\rm  SUSY}^{\rm Higgs}$ in~(\ref{ee6}),  which
entirely  originates   from  Higgs-boson exchange   diagrams  in   the
CP-violating    MSSM.     Based  on    our  resummed   FCNC  effective
Lagrangian~(\ref{master}), we    obtain  in   the  zero   strong-phase
approximation
\begin{eqnarray}
  \label{direct}
\Omega_{\rm SUSY}^{\rm Higgs} \!\!&=&\!\! 2\, \sum_{i=1}^3\, \sum_{q=u,d,s}\,
\frac{m_s\, m_q}{M_{H_i}^2} \nonumber\\[3mm]
&&\hspace{-1,3cm}
\times\, \Bigg[\, 
\frac{\Im m(g_{H_i\bar{s}d}^L)\, g_{H_i\bar{q}q}^S}{A^2\lambda^5\eta}\,
\Bigg(\, \frac{\bra{\pi^+\pi^-}(\bar{s}P_L d )(\bar{q}q) \ket{K^0}_0}{y_6\, 
\bra{\pi^+\pi^-} Q_6 \ket{K^0}_0}\ -\ \frac{1}{|\omega|}\, 
\frac{\bra{\pi^+\pi^-}(\bar{s}P_L d )(\bar{q}q) \ket{K^0}_2}{y_6\, 
\bra{\pi^+\pi^-} Q_6 \ket{K^0}_0}\, \Bigg)\nonumber \\[3mm]
&&\hspace{-1.3cm}
+\, \frac{\Re e (g_{H_i\bar{s}d}^L)\, g_{H_i\bar{q}q}^P}{A^2\lambda^5\eta}\,
\Bigg(\, \frac{\bra{\pi^+\pi^-}(\bar{s}P_L d )(\bar{q}\gamma_5q)
 \ket{K^0}_0}{y_6\, \bra{\pi^+\pi^-} Q_6 \ket{K^0}_0}\ -\ \frac{1}{|\omega|}\,
\frac{\bra{\pi^+\pi^-}(\bar{s}P_L d )(\bar{q}\gamma_5 q) \ket{K^0}_2}{y_6\, 
\bra{\pi^+\pi^-} Q_6 \ket{K^0}_0}\, \Bigg)\, \Bigg]\nonumber\\[3mm]
&&\hspace{-1.3cm} +\, 2 \sum_{i=1}^3
\sum_{q=u,d,s}\, \frac{m_d\, m_q}{M_{H_i}^2}\\[3mm] 
&&\hspace{-1.3cm}
\times\, \Bigg[\, \frac{\Im m(g_{H_i\bar{s}d}^R)\, 
g_{H_i\bar{q}q}^S}{A^2\lambda^5\eta}\, 
\Bigg(\, \frac{\bra{\pi^+\pi^-}(\bar{s}P_R d )(\bar{q}q) \ket{K^0}_0}{y_6\, 
\bra{\pi^+\pi^-} Q_6 \ket{K^0}_0}\ -\ \frac{1}{|\omega|}\, 
\frac{\bra{\pi^+\pi^-}(\bar{s}P_R d )(\bar{q}q) \ket{K^0}_2}{y_6\, 
\bra{\pi^+\pi^-} Q_6 \ket{K^0}_0}\, \Bigg)\nonumber\\[3mm]
&&\hspace{-1.3cm} +\,
\frac{\Re e (g_{H_i\bar{s}d}^R)\, g_{H_i\bar{q}q}^P}{A^2\lambda^5\eta}\,
\Bigg(\, \frac{\bra{\pi^+\pi^-}(\bar{s}P_R d )(\bar{q}\gamma_5q)
 \ket{K^0}_0}{y_6\, 
\bra{\pi^+\pi^-} Q_6 \ket{K^0}_0}\ -\ \frac{1}{|\omega|}\, 
\frac{\bra{\pi^+\pi^-}(\bar{s}P_R d )(\bar{q}\gamma_5 q) \ket{K^0}_2}{y_6\, 
\bra{\pi^+\pi^-} Q_6 \ket{K^0}_0}\, \Bigg)\, \Bigg]\, ,\nonumber
\end{eqnarray}
where the  subscripts 0,  2 adhered  to  the hadronic  matrix elements
denote the total isospin  $I$ of the  final states, and $A^2 \lambda^5
\eta$   is   a  CKM  combination in the Wolfenstein  parameterization,
which has the value $A^2 \lambda^5  \eta \approx 1.3\times 10^{-4}$ in
the  SM.   Furthermore, for $\Lambda_{\rm  QCD}  = 325$~MeV and $m_s =
150$ MeV, the SM  Wilson coefficient $y_6$  and the matrix element  of
$Q_6$    take  on   the values~\cite{Bosch}:    $y_6  =  -0.089$   and
$\bra{\pi^+\pi^-} Q_6 \ket{K^0}_0 = -0.35 {\rm ~GeV}^3$, respectively.
Also,   experimental analyses suggest   the  value $|\omega| = 0.045$,
approximately yielding the SM contribution  $\Omega_{\rm SM}\sim 1$ to
$\epsilon'/\epsilon$  in~(\ref{ee6}).   Finally,    the     parameters
$g^{S,P}_{H_i\bar{d}d}$      and  $g^{S,P}_{H_i\bar{u}u}$ that   occur
in~(\ref{direct}) are  the diagonal scalar and  pseudoscalar couplings
of  the $H_i$  bosons to $u$-  and  $d$-type quarks~\cite{CEPW}, whose
strengths  are normalized  to  the  SM  Higgs-boson coupling.    These
reduced $H_i$-couplings are given by
\begin{eqnarray}
  \label{HSPd}
g^S_{H_i\bar{d}d} \!&=&\! {\textstyle \frac{1}{2}}\, \big(\, 
g^L_{H_i\bar{d}d}\: +\:  g^R_{H_i\bar{d}d}\,\big)\,,\qquad
g^P_{H_i\bar{d}d} \  =\ {\textstyle \frac{i}{2}}\, \big(\, 
g^L_{H_i\bar{d}d}\: -\:  g^R_{H_i\bar{d}d}\,\big)\,,\qquad\\
  \label{HSPu}
g^S_{H_i\bar{u}u} \!&=&\! O_{2i}/\sin\beta\,, \qquad
g^P_{H_i\bar{u}u} \  =\ -O_{3i}\cot\beta\,,
\end{eqnarray}
where we  have neglected the  small radiative threshold effects in the
up sector.

On the experimental side, the  latest world average  result for $\Re e
(\epsilon'/\epsilon)$ is~\cite{Nir} 
\begin{equation}
  \label{eprimexp}
\Re e\bigg(\, \frac{\epsilon'}{\epsilon}\,\bigg)\ =\  (1.66 \pm
0.16)\times 10^{-3}\; ,
\end{equation}
at  the  1-$\sigma$  confidence  level  (CL).  In  the  light  of  the
experimental result~(\ref{eprimexp})  and the discussion  given above,
we may conservatively require that
\begin{eqnarray}
  \label{boundepe}
|\Omega_{\rm SUSY}^{\rm Higgs}|\ \stackrel{<}{{}_\sim}\ 1 \;.
\end{eqnarray}
The   biggest contribution in  the  sum  over quarks in~(\ref{direct})
comes   from the  $d$-quark   and  exhibits  the qualitative  scaling
behaviour
\begin{eqnarray}
  \label{eprimapprox}
\Omega_{\rm SUSY}^{\rm Higgs}\ \simeq\ \frac{2 m_s m_d}{M_H^2} \,
\frac{\tan^3\beta}{|\omega|}\, \times\, {\cal O}(1)\;.
\end{eqnarray}
For    instance,    for     $\tan\beta=50$    and     $M_H=200$   GeV,
(\ref{eprimapprox})  gives $\Omega_{\rm SUSY}^{\rm  Higgs}  \simeq 0.1
\times {\cal O}(1)$.  Obviously, such a contribution is, in principle,
non-negligible, but very sensitively  depends on the actual values  of
the new hadronic matrix elements:
\begin{eqnarray}
  \label{QLRSP}
(Q^{L,R}_{S,P})_I\ =\ \bra{\pi^+\pi^-}\,
(\bar{s}P_{L,R} d )\,(\bar{q}(1,\gamma_5) q)\, \ket{K^0}_I \;,
\end{eqnarray}
with $I=0,2$.  A detailed calculation of  the hadronic matrix elements
$(Q^{L,R}_{S,P})_I$  in~(\ref{QLRSP})  will be   given elsewhere.   In
Section~4, however, we will  present  numerical estimates of   $\Delta
M_K$ and $\epsilon_K$ within the context of generic soft SUSY-breaking
models.

\subsection{$\Delta M_{B_q}$, $B_q\to \ell^+\ell^-$ and associated 
  CP asymmetries}

We start our discussion of a set of observables in the $B$-meson
system by first analyzing the $B^0_q$-$\bar{B}^0_q$ mass difference,
$\DmBq$ with $q={s,d}$, in the CP-violating MSSM at large $\tan\beta$.
In the applicable limit of equal $B$-meson lifetimes, $\DmBq$ may be
written as the modulus of a sum of a SM and a SUSY term:
\begin{eqnarray} 
\DmBq\ =\ 2\, 
| \bra{\bar{B}_q^0}\, H_{\rm eff}^{\Delta B=2}\, \ket{B^0_q}_{\rm SM}\ + \
\bra{\bar{B}_q^0}\, H_{\rm eff}^{\Delta B=2}\, \ket{B^0_q}_{\rm SUSY}|\;,
\end{eqnarray}
where  the effective  $\Delta  B=2$  Hamiltonian $H_{\rm  eff}^{\Delta
B=2}$ may be obtained from the $\Delta S=2$ one stated in~(\ref{DS2}),
after  making the  obvious replacements:  $s\to b$ and  $d\to q$, with
$q=d,s$.  Proceeding as in Section~3.1, we  arrive at analogous closed
expressions for the SUSY  contributions to the $\Delta B=2$ transition
amplitudes:
\begin{eqnarray}
  \label{Bsusy}
\bra{\bar{B}^0_d}\,H_{\rm eff}^{\Delta B=2}\,
\ket{B^0_d}_{\rm SUSY} \!\!&=&\!\!  1711 {\rm ~ps}^{-1} 
\Bigg(\,\frac{\hat{B}_{B_d}^{1/2}\, F_{B_d}}{230 {\rm ~MeV}}\, \Bigg)^2\,
\Bigg(\,\frac{\eta_B}{0.55}\,\Bigg) \nonumber\\[3mm] 
&&\hspace{-2.7cm}
\times\, \Big[\, 0.88\, \Big(\,
C_2^{\rm  LR\, (DP)}\: +\: C_2^{\rm  LR\, (2HDM)}\, \Big)\ -\
0.52\, \Big(\, C_1^{\rm  SLL\,(DP)}\: +\: 
C_1^{\rm  SRR\,(DP)}\, \Big)\, \Big]\;, \nonumber \\[3mm]
\bra{\bar{B}^0_s}\, H_{\rm eff}^{\Delta B=2}\, \ket{B^0_s}_{\rm SUSY} 
\!\!&=&\!\! 
2310 {\rm ~ps}^{-1} \Bigg(\,
\frac{\hat{B}_{B_s}^{1/2}\, F_{B_s}}{265 {\rm ~MeV}}\, \Bigg)^2
\Bigg(\,\frac{\eta_B}{0.55}\,\Bigg) \nonumber\\[3mm] 
&&\hspace{-2.7cm}
\times\, \Big[\, 0.88\, \Big(\,
C_2^{\rm  LR\, (DP)}\: +\: C_2^{\rm  LR\, (2HDM)}\, \Big)\ -\
0.52\, \Big(\, C_1^{\rm  SLL\,(DP)}\: +\: 
C_1^{\rm  SRR\,(DP)}\, \Big)\, \Big]\;.\qquad\quad
\end{eqnarray}
In deriving~(\ref{Bsusy}),  we   have  also  substituted the    values
determined   in~\cite{Jager,Ciuchini,Misiak,Becirevic,Chankowski}  for
the NLO-QCD factors, along with their hadronic  matrix elements at the
scale $\mu=4.2$~GeV:
\begin{eqnarray}
\bar{P}_1^{\rm LR}\ =\ -0.58 \;,\qquad
\bar{P}_2^{\rm LR}\ =\ 0.88 \;,\qquad
\bar{P}_1^{\rm SLL}\ =\ -0.52 \;,\qquad
\bar{P}_2^{\rm SLL}\ =\ -1.1 \;.
\end{eqnarray}
Moreover,  the   corresponding Wilson   coefficients         appearing
in~(\ref{Bsusy})  may  be recovered  from those  in~(\ref{dpkaon}) and
(\ref{2HDM}), after performing the quark replacements mentioned above.

Another observable, which is   enhanced at large $\tan\beta$,   is the
pure            leptonic              decay         of             $B$
mesons~\cite{FCNC,BK,FCNCCP,Frank,Isidori,Dedes1,ADKT,Ambrosio,Buras,MTW},
$\bar{B}^0_q   \to \ell^+  \ell^-$,   with   $\ell  = \mu$,    $\tau$.
Neglecting contributions  proportional  to the   lighter quark  masses
$m_{d,s}$, the  relevant effective Hamiltonian  for  $\Delta B=1$ FCNC
transitions, such as $b\to  q \ell^+ \ell^-$  with $q = d,s$, is given
by
\begin{eqnarray}
  \label{DB1}
H_{\rm eff}^{\Delta B=1}\ =\ -\,2\,\sqrt{2}\, G_F\, V_{tb}V_{tq}^*\,
\Big(\, C_S\, {\cal O}_S\ +\ C_P\, {\cal O}_P\ +\ C_{10}\, {\cal O}_{10}
\Big)\;,
\end{eqnarray}
where 
\begin{eqnarray}
{\cal O}_S &=& \frac{e^2}{16\pi^2}\, m_b\, 
(\bar{q} P_R b)\, (\bar{\ell}\ell) \;, \nonumber \\[2mm]
{\cal O}_P &=& \frac{e^2}{16\pi^2}\, m_b\, (\bar{q} P_R b)\, 
(\bar{\ell}\gamma_5 \ell) \;,\nonumber \\[2mm]
{\cal O}_{10} &=& \frac{e^2}{16\pi^2}\,  (\bar{q}\gamma^\mu P_L b)\,
 (\bar{\ell}\gamma_\mu \gamma_5 \ell) \;.
\end{eqnarray}
Employing our resummed FCNC effective Lagrangian~(\ref{master}), it is
not difficult  to compute the Wilson  coefficients $C_S$ and  $C_P$ in
the region  of  large values of $\tan\beta$:\footnote{Our  approach to
Higgs-mediated FCNC  effects    presented here  may   be  extended  to
consistently account for charged-lepton flavour violation in $B$-meson
decays, such  as  $B_{s,d}\to \ell^+ \ell'^{-}$~\cite{DER}, where  the
effective     off-diagonal        Higgs-lepton-lepton        couplings
$g_{H_i\bar{\ell}\ell'}^{S,P}$    can    be derived    by  following a
methodology very analogous to the one described in Section~2.}
\begin{eqnarray}
  \label{CSCP}
C_S &=& \frac{2 \pi m_\ell}{\alpha_{\rm em}}\, 
\frac{1}{V_{tb} V_{tq}^*}\, \sum_{i=1}^3\, \frac{g_{H_i\bar{q}b}^R\, 
g_{H_i\bar{\ell}\ell}^S}{M_{H_i}^2} \ ,\nonumber\\[3mm]
C_P &=& i\, \frac{2 \pi m_\ell}{\alpha_{\rm em}}\, \frac{1}{V_{tb} V_{tq}^*}\,
\sum_{i=1}^3\, \frac{g_{H_i\bar{q}b}^R\, 
g_{H_i\bar{\ell}\ell}^P}{M_{H_i}^2} \ ,
\end{eqnarray}
while  $C_{10}=-4.221$ is the   leading SM  contribution.  In  analogy
to~(\ref{HSPd}), the reduced scalar  and pseudoscalar  Higgs couplings
to charged  leptons $g_{H_i\bar{\ell}\ell}^{S,P}$  in~(\ref{CSCP}) are
given by
\begin{equation}
g_{H_i \bar{\ell} \ell}^S\ =\ \frac{O_{1i}}{\cos\beta}\ , \qquad
g_{H_i \bar{\ell} \ell}^P\ =\ -\,\tan\beta\, O_{3i} \; ,
\end{equation}
where non-holomorphic vertex effects on the  leptonic sector have been
omitted  as  being   negligibly   small.

With the approximations mentioned  above, the branching ratio  for the
$\bar{B}^0_q$  meson decay  to   $\ell^+ \ell^-$  acquires the  simple
form~\cite{Frank}
\begin{eqnarray}
  \label{Bll}
{\cal B}(\bar{B}^0_q \to \ell^+ \ell^-) &=& \\
&&\hspace{-2cm}
\frac{G_F^2 \alpha_{\rm em}^2}{16\pi^3}\, M_{B_q} \tau_{B_q}\, 
|V_{tb}V_{tq}^*|^2\, \sqrt{1-\frac{4 m_\ell^2}{M_{B_q}^2}}\
\Bigg[\, \Bigg(\,1-\frac{4 m_\ell^2}{M_{B_q}^2} \Bigg)\, |F^q_S|^2\ +\
|F_P^q\: +\: 2 m_\ell F_A^q|^2\, \Bigg] \;,\nonumber
\end{eqnarray}
where $\tau_{B_q}$ is the total lifetime of the $B_q$ meson and
\begin{eqnarray}
  \label{FSP}
F_{S,P}^q\ =\ -\,\frac{i}{2}\, M_{B_q}^2 F_{B_q}\, 
\frac{m_b}{m_b+m_q}\, C_{S,P}\;,\qquad 
F_A^q\ =\ -\,\frac{i}{2}\,F_{B_q}\, C_{10} \;.
\end{eqnarray}
In our numerical estimates,  we ignore the contribution from $C_{10}$,
as being  subdominant in  the region of  large $\tan\beta$,  i.e.\ for
$\tan\beta \stackrel{>}{{}_\sim} 40$,  where all Higgs-particle masses
are  well below  the TeV  scale.  The SM  predictions as  well as  the
current  experimental bounds  pertinent to  ${\cal  B}(\bar{B}^0_d \to
\ell^+ \ell^-)$ can be read off from Table~1 in~\cite{Dedes2}.
  
In the CP-violating  MSSM, an equally  important class of  observables
related to ${\cal B}(\bar{B}^0_{d,s}  \to \ell^+ \ell^-)$~\cite{IN} is
the  one probing possible  CP asymmetries  that can take  place in the
same leptonic $B$-meson decays.  The leptonic  CP asymmetries may shed
even light on the CP nature of possible new-physics effects, as the SM
prediction for these   observables turns out to  be  dismally small of
order $10^{-3}$~\cite{Liao}.   This SM result is  a consequence of the
fact that the CP-violating phase in $B^0$-$\bar{B}^0$-mixing parameter
$q/p$  is opposite to   the one entering the  ratio  of the amplitudes
$\bar{{\cal        A}}_{L(R)}(\bar{B}^0_{d,s}      \to      l^+_{L(R)}
l^-_{L(R)})/{\cal A}_{L(R)} ({B}^0_{d,s} \to l^+_{L(R)}  l^-_{L(R)})$,
such  that the net  CP-violating   effect on the observable  parameter
$\lambda_{L(R)}  = (q/p)    (\bar{{\cal  A}}_{L(R)}/{\cal  A}_{L(R)})$
almost cancels out.
 
There are two  possible time-dependent CP asymmetries associated  with
leptonic $B$-meson decays that are physically allowed:
\begin{eqnarray}
  \label{ACPlept}
{\cal A}_{\rm CP}^{(B_q^0\to l_L^+ l_L^-)}\ =\
\frac{\int_0^{\infty} dt \, \Gamma(B_q^0(t)\to l_L^+l_L^-)\: 
-\: \int_0^{\infty}  dt \, \Gamma(\bar{B}_q^0(t)\to l_R^+l_R^-)}
{\int_0^{\infty} dt \, \Gamma(B_q^0(t)\to l_L^+l_L^-)\: 
+\: \int_0^{\infty} dt \, \Gamma(\bar{B}_q^0(t)\to l_R^+l_R^-)} 
\end{eqnarray}
and  ${\cal   A}_{\rm    CP}^{(B_q^0 \to   l_R^+   l_R^-)}$,  with  $L
\leftrightarrow R$.  Under the assumption  that $q/p$ is a pure phase,
one finds~\cite{Liao}
\begin{eqnarray}
 \label{bmumuass}
{\cal   A}_{\rm CP}^{(B_q^0\to  l_L^+ l_L^-)}\  =\ -\,\frac{2 x_q\,\Im m
\lambda_q} {(2\, +\, x_q^2)\: +\:  x_q^2\, |\lambda_q|^2}\ ,\qquad
{\cal A}_{\rm  CP}^{(B_q^0\to l_R^+ l_R^-)}\ =\ -\, \frac{2  
x_q\,\Im m \lambda_q}{(2\,+\,x_q^2)\,|\lambda_q|^2\: +\: x_q^2 } \ ,\quad
\end{eqnarray} 
where $x_q = \Delta M_{B_q}/\Gamma_{B_q}$ and
\begin{eqnarray}
  \label{lq}
\lambda_q\ =\ \frac{M^{q*}_{12}}{|M^q_{12}|}\ 
\Bigg(\,\frac{V_{tb}V_{tq}^*}{V^*_{tb}V_{tq}}\,\Bigg)\
\frac{\beta_l\,C_S \: +\: C_P\: +\: 2m_l C_{10}/(m_b M_{B_q})}{
\beta_l\,C_S^*\: -\: C_P^*\: -\: 2m_l C_{10}/(m_b M_{B_q})} \ .
\end{eqnarray}
In (\ref{lq}), $\beta_l =  (1-4 m_l^2/M_{B_q}^2)^{1/2}$, $M^q_{12}$ is
the dispersive part  of the $B^0_q$-$\bar{B}^0_q$ matrix  element, and
$C_{S,P}$ are Wilson  coefficients given in~(\ref{CSCP}).  The maximal
value that the  leptonic CP asymmetries in~(\ref{bmumuass}) can  reach
is ${\cal A}_{\rm CP }^{\rm max}=1/\sqrt{2+x_q^2}$ and is obtained for
$\Im    m\lambda_q  = |\lambda_q  |^2$.    {}From current experimental
data~\cite{PDG},  one  may    extract    the values  $x_d=0.76$    and
$x_s\stackrel{>}{{}_\sim} 19$ at the 95\%   CL, which leads to  ${\cal
A}_{\rm CP }^{\rm max}(B_s) \approx 5\%$ and  ${\cal A}_{\rm CP }^{\rm
max}(B_d) \approx 62\%$.

Within the   framework of our resummed  FCNC  effective Lagrangian, we
also improve earlier calculations~\cite{Liao} of the CP asymmetries by
including   $B^0_q$-$\bar{B}^0_q$ mixing       effects         through
$M^{q*}_{12}/|M^q_{12}|$  in~(\ref{lq}).   According  to our  standard
approach of splitting  the amplitude  into a SM  and  a MSSM part,  we
obtain for the SM part
\begin{eqnarray}
M_{12}^{q*\, ({\rm SM})}\ =\ \frac{G_F^2 M_W^2}{12\pi^2}\ M_{B_q} \eta_B 
\hat{B}_{B_q} F_{B_q}^2\, (V_{tq}V_{tb}^*)^2\, S_{tt} \;,
\end{eqnarray}
where $S_{tt}\approx  2.38$  is     the   value  of   the     dominant
$m_t$-dependent loop function  for a  top-pole mass  $m_t =  175$ GeV.
The    SUSY   contribution   to     $M_{12}^{q*\,   ({\rm  SUSY})}   =
\bra{\bar{B}_d^0}\, H_{\rm eff}^{\Delta B=2}\, \ket{B^0_d}_{\rm SUSY}$
may be obtained from (\ref{Bsusy}). 

In the next section, we will present numerical  estimates for the $K$-
and  $B$-meson FCNC observables,  based  on  the analytic  expressions
derived above.

\setcounter{equation}{0} 
\section{Numerical estimates}

In  this section,  we shall  numerically   analyze the  impact of  the
$\tan^2\beta$-enhanced  FCNC  interactions  on  a number  of  $K$- and
$B$-meson  observables which were discussed in  detail in Sections 3.1
and 3.2, such  as  $\DmK$, $\epsilon_K$,  $\DmBd$, $\DmBs$, $B_d   \to
\tau^+ \tau^-$, $B_s \to \mu^+ \mu^-$ and their associated leptonic CP
asymmetries.   For   our  illustrations,  we   consider   two  generic
low-energy soft SUSY-breaking scenarios, (A) and (B).

In scenario~(A), the squark masses are taken to be universal and $\Eg$
and  $\Eu$   are  proportional  to  the  unity    matrix at the   soft
SUSY-breaking scale $M_{\rm SUSY}$.  The CP-conserving version of this
scenario has frequently  been  discussed in the literature  within the
context of minimal flavour-violation models, see e.g.~\cite{BK}.

In scenario~(B)  we assume the existence of  a  mass hierarchy between
the first two generations of squarks and  the third generation, namely
the first two generations are degenerate and can  be much heavier than
the  third  one.  In addition,  although  not mandatory, we assume for
simplicity  that     the  model-dependent   unitary    matrix   $\ULQ$
in~(\ref{Epars})  is such  that  $\Egb$   and $\Eub$ become   diagonal
matrices in this scenario. Clearly, in  the limit in which all squarks
are degenerate, scenario~(B) coincides with (A).

In Fig.~\ref{fig1} we give  a schematic representation of  the generic
mass  spectrum that will  be assumed in  our numerical analysis.  More
explicitly,  we fix  the charged Higgs-boson   $M_{H^+}$ to the  value
$200$~GeV.   Since       the  effect   of     the     gaugino-Higgsino
mixing~\cite{MTW,IN}  on   the resummation  matrix   ${\bf R}$  can be
significantly   reduced for $m_{\tilde{W}}  \ll   \mu$, we ignore this
contribution by considering  the  relatively low  value $m_{\tilde{W}}
\approx 2m_t \ll M_{\rm SUSY}$ in our numerical estimates, with $m_t =
175$~GeV. As can  be  seen from Fig.~\ref{fig1},  the third-generation
soft squark mass $m_{\tilde{t}}$, the $\mu$-parameter, the gluino mass
$m_{\tilde{g}}$ and the trilinear soft  Yukawa coupling $A_U$ are set,
for simplicity reasons, to the common soft SUSY-breaking scale $M_{\rm
SUSY}$, which is typically taken to be 1~TeV.

The soft squark masses of the other two  generations are assumed to be
equal to $m_{\tilde{q}}$  in our generic framework.   To account for a
possible  hierarchical   difference   between   the     mass    scales
$m_{\tilde{t}}$ and   $m_{\tilde{q}}$,   we introduce  the   so-called
hierarchy factor $\rho$, such that $m_{\tilde{q}} = \rho m_{\tilde{t}}
= \rho M_{\rm SUSY}$.  As has been mentioned  above, models of minimal
flavour violation  correspond to scenario~(A) with $\rho  = 1$.  As we
will see in detail in  Sections 4.1 and  4.2, the predictions for  the
$K$- and $B$-meson FCNC observables crucially depend  on the values of
the  hierarchy factor $\rho$.   Equally important modifications in the
predictions are obtained for different values of the soft CP-violating
phases $\phi_g  = {\rm arg}\,  (m_{\tilde{g}})$ and $\phi_{A_U} = {\rm
arg}\, (A_U)$.  In addition,  the $K$- and $B$-meson FCNC  observables
exhibit a non-trivial dependence on  the CKM phase $\delta_{\rm CKM}$,
which is  varied independently in our  figures.  

Although we primarily use $\tan\beta = 50$  and $M_{H^+} = 0.2$~TeV as
inputs in our   numerical analysis, approximate predictions  for other
values of  the   input parameters may be   obtained  by  rescaling the
numerical estimates by a factor
\begin{equation}
  \label{scale}  
x_{\cal O}\ =\ \bigg(\,\frac{\tan\beta}{50}\,\bigg)^n\:
\times\: \bigg(\,\frac{0.2~{\rm TeV}}{M_{H^+}}\,\bigg)^k\,,
\end{equation}
where the integers  $n$ and $k$ depend  on the FCNC  observable ${\cal
O}$ under  study.  Such a rescaling proves  to be  fairly accurate for
$\tan\beta    \stackrel{>}{{}_\sim}          40$      and     $M_{H^+}
\stackrel{>}{{}_\sim} 150$~GeV, which  is the kinematic  region of our
interest.

\subsection{$\DmK$ and $|\epsilon_K|$}

The SM effects on $\DmK$ and $|\epsilon_K|$ were extensively discussed
in the literature~\cite{NH}, so we will not dwell upon this issue here
as well. Instead, we assume that the SM explains well the experimental
results for the above two observables~\cite{PDG}:
\begin{eqnarray}
  \label{expDmK}
\Delta M_K^{\rm exp} &=& (3.490 \pm 0.006)\times 10^{-12} ~{\rm MeV}\,, \\
  \label{expek}
|\epsilon^{\rm exp}_K| &=& (2.282 \pm 0.017)\times 10^{-3}\;.
\end{eqnarray}
Given  the significant  uncertainties in  the calculation of  hadronic
matrix elements, however,  our approach will be  to constrain the soft
SUSY-breaking   parameters by  conservatively   requiring that $\Delta
M_K^{\rm  SUSY}$ and $|\epsilon_K^{\rm SUSY}|$ do  not  exceed in size
the SM   predictions.\footnote{Both  $\DmK$ and  $|\epsilon_K|$  place
important constraints on   the  $\rho$-$\eta$ plane of the   unitarity
triangle.  The so-derived limits can be used to constrain new physics.
In this case, a global fit of all the relevant FCNC observables to the
unitarity triangle might  be more appropriate.   We intend to  address
this issue in a future~work.}

\begin{figure}[p]
\hspace*{3.5cm}\psfig{figure=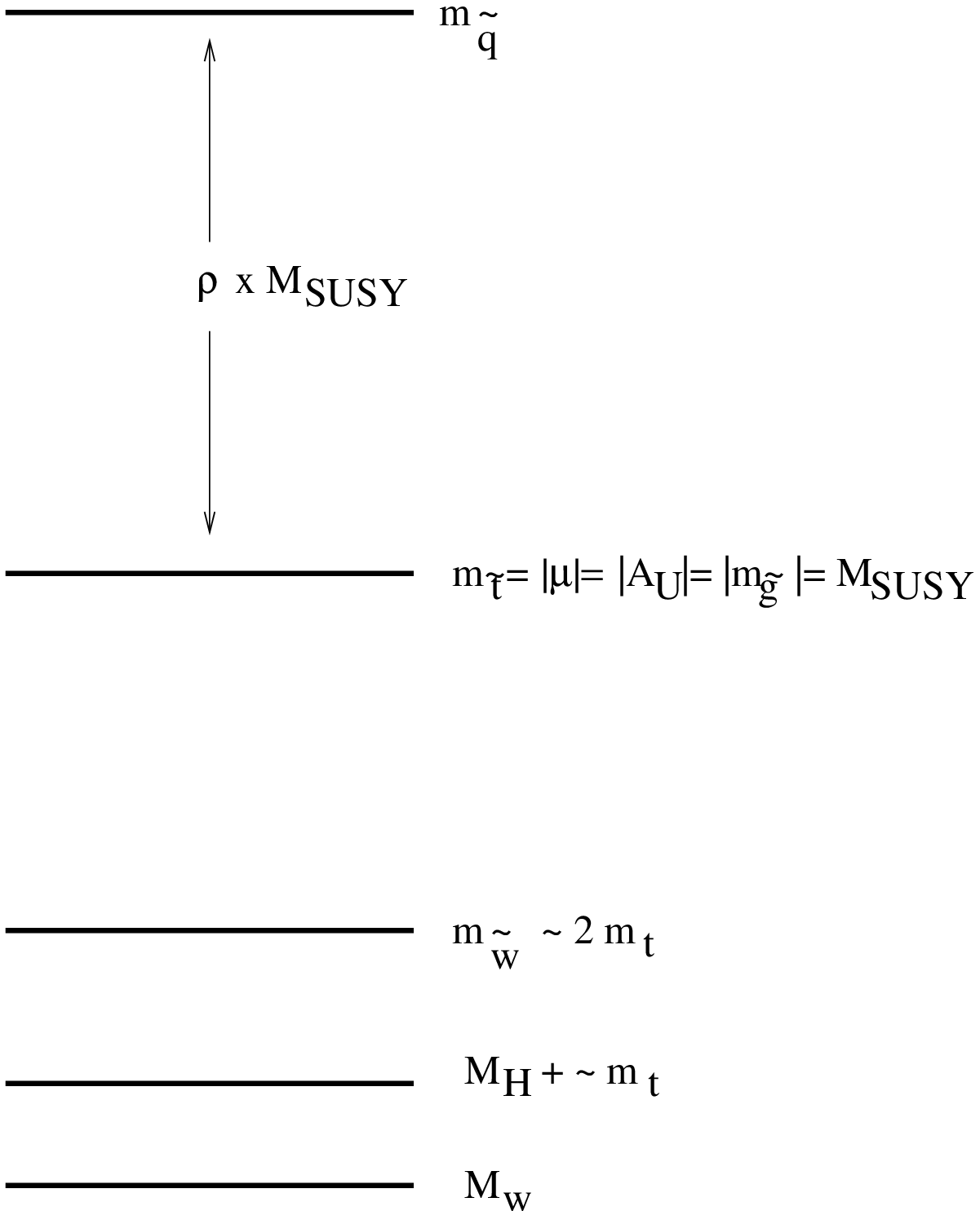}
\caption{{\it  Schematic  representation  of the   SUSY mass  spectrum
considered in  our    numerical analysis,  where  $m_{\tilde  q}$  and
$m_{\tilde  t}$  denote   the  masses of   the  first   two  and third
generations of   squarks, respectively.  The  hierarchy factor $\rho$,
the   phase $\phi_{\tilde{g}}$ of   the   gluino mass, and  the phases
$\phi_{A_{t,b}}$ of the soft SUSY  breaking trilinear couplings,  with
$\phi_{A_t} = \phi_{A_b}  = \phi_{A_U}$, are varied independently (see
also discussion in the text).}}\label{fig1}
\end{figure}

\begin{figure}[p]
\psfig{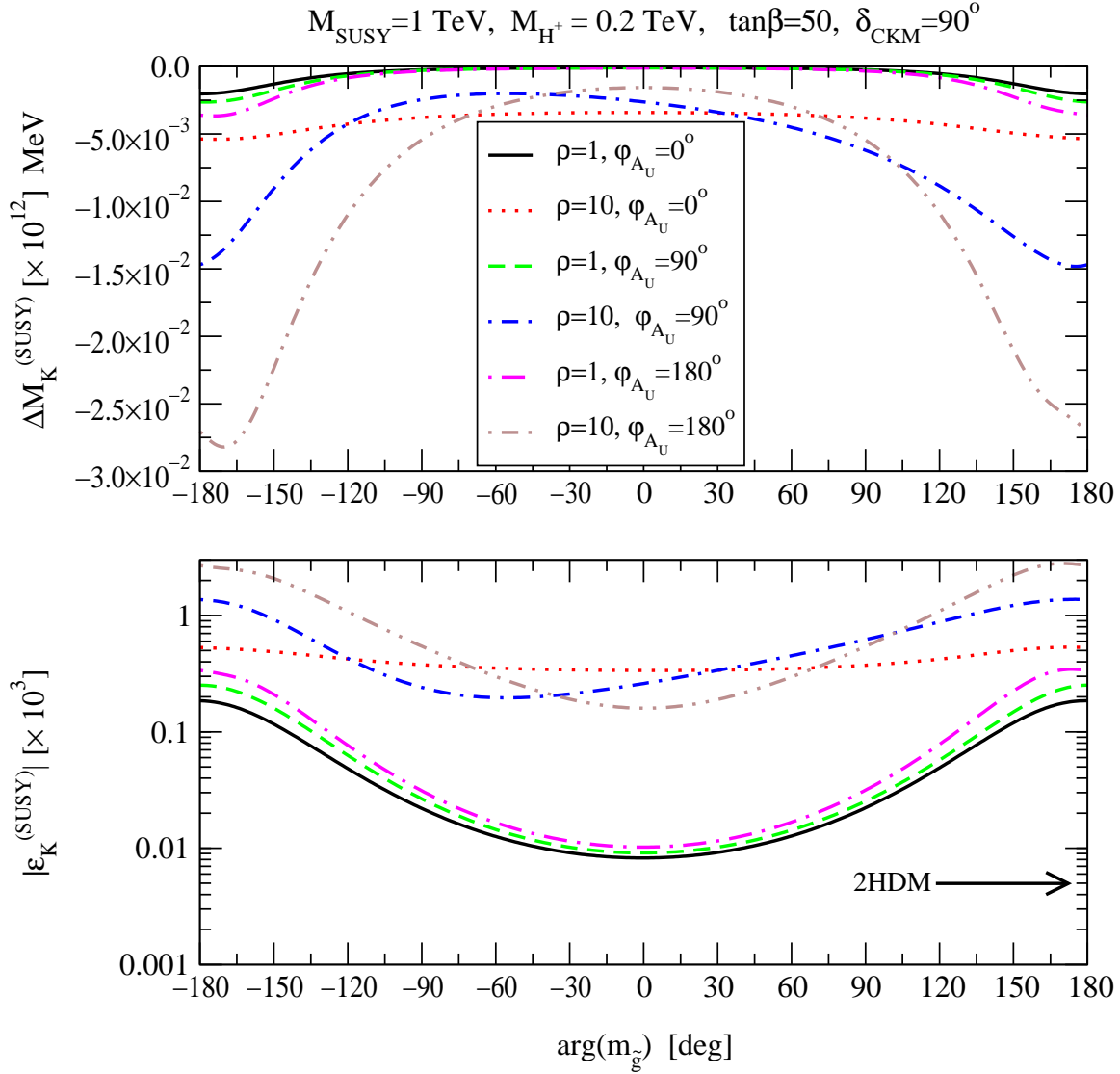}
\caption{{\it SUSY Higgs-DP contributions  to $\epsilon_K$ and $\Delta
M_K$ given in units of $10^{-12}$ MeV  and $10^{-3}$, respectively, as
functions   of the gluino   phase  ${\rm arg}\,  (m_{\tilde{g}})$, for
$M_{\rm   SUSY} = 1$~TeV,  $M_{H^+}  = 0.2$~TeV, $\tan\beta  = 50$ and
$\delta_{\rm CKM}  =  90^\circ$.   As  is  shown  above, the different
curves  are  obtained      for   selected  values  of  $\rho$      and
$\phi_{A_U}$. The  size of the  2HDM  effect alone on  $\epsilon_K$ is
indicated by an arrow.}}\label{fig2}
\end{figure}

To start with, we display  in Fig.~\ref{fig2} numerical values for the
Higgs-boson DP effects  on $\DmK$ and  $|\epsilon_K|$ as  functions of
the   gluino phase ${\rm  arg}\,  m_{\tilde{g}}$,  where the hierarchy
factor  $\rho$ and the phase $\phi_{A_U}  = {\rm  arg}\, (A_{t,b})$ of
the soft SUSY-breaking  trilinear Yukawa couplings assume the discrete
values:  $(\rho,\ \phi_{A_U}) = (1,\  0^\circ),\ (10,\ 0^\circ),\ (1,\
90^\circ),\  (10,\ 90^\circ),  (1,\  180^\circ),\  (10,\  180^\circ)$.
According to our  CP-phase  conventions~\cite{CEPW}, $\mu$   is always
taken to be positive, while the CKM phase $\delta_{\rm CKM}$ is chosen
to its   maximal value $90^\circ$.     The subdominant  one-loop  2HDM
contribution   coming     from $W^\pm$-$H^\mp$   box     graphs  [cf.\
(\ref{2HDM})] has also been indicated by  an arrow in Fig.~\ref{fig2}.
Predictions for $M_{H^+}$    and $\tan\beta$ values other   than those
shown in Fig.~\ref{fig1} may be  approximately obtained by multiplying
the   numerical   estimates    by    a     factor $x_{\cal   O}      =
(\tan\beta/50)^4\times  (0.2~{\rm  TeV}/M_{H^+})^2$.    We observe  in
Fig.~\ref{fig2} that the  resulting values for $\Delta M_K^{\rm SUSY}$
can exceed  the experimental error in (\ref{expDmK})  by  one order of
magnitude,  for    $\rho=10$  and  $|\phi_{A_U}|,\  |\phi_{\tilde{g}}|
\stackrel{>}{{}_\sim}   90^\circ$.       For   the     same    inputs,
$|\epsilon_K^{\rm    SUSY}|$ takes   on   values   comparable to   the
experimentally measured one (\ref{expek}).  Here, we should stress the
fact that universal squark-mass scenarios corresponding  to $\rho = 1$
can  still  predict sizeable  effects  on $\epsilon_K$.  This non-zero
result should  be contrasted  with  the one  of the  gluino-squark box
contributions  to   $\DmK$  and  $|\epsilon_K|$~\cite{Nilles} which do
vanish   in the   limit  of strictly    degenerate squarks   due to  a
SUSY-GIM-cancellation mechanism.

In our case of Higgs-mediated FCNC observables, however, the situation
is slightly different.  As we have discussed in Section~2, the size of
the FCNC  effects is  encoded in the  flavour  structure of the 3-by-3
resummation matrix  $\Ro$.  Since $\Ro$  is diagonal for the scenarios
(A)  and     (B)     under consideration,   we    can     expand   the
$\tan\beta$-enhanced FCNC terms    $(\Vckm^\dagger\,        \Ro^{-1}\,
\Vckm)_{dd'}$ in~(\ref{master}) as follows:
\begin{eqnarray}
  \label{VRV}
(\Vckm^\dagger\, \Ro^{-1}\, \Vckm)_{dd'}\  =\ V^*_{ud}\, R^{-1}_u\, V_{ud'}\:
+\: V^*_{cd}\, R^{-1}_c\, V_{cd'}\: +\: V^*_{td}\, R^{-1}_t\, V_{td'} \;,
\end{eqnarray}
where  $d$  and $d'$ collectively denote    all down-type quarks, with
$d\neq d'$.    For  the   parameters  adopted  in~Fig.~\ref{fig1}, the
quantities $R^{-1}_{u,c,t}$  can be simplified further to\footnote{For
$\rho\stackrel{>}{{}_\sim}         10$,   the  first   two   equations
in~(\ref{express})  may be better  approximated by replacing $\rho \to
\rho/\sqrt{2}$.}
\begin{eqnarray}
  \label{express}
R^{-1}_u &\approx& \bigg[\, 1\: +\: \bigg(\, 
\frac{\alpha_s}{3\pi\rho^2}\, e^{-i\phi_{\tilde{g}}}\: +\: 
\frac{|h_u|^2}{32\pi^2\rho^2}\, 
e^{-i\phi_{A_U}}\,\bigg)\tan\beta\, \bigg]^{-1}\;, \nonumber \\[3mm]
R^{-1}_c &\approx& \bigg[\, 1\: +\: \bigg(\,
\frac{\alpha_s}{3\pi\rho^2}\, e^{-i\phi_{\tilde{g}}}\: +\: 
\frac{|h_c|^2}{32\pi^2\rho^2}\, e^{-i\phi_{A_U}}\,\bigg)\,\tan\beta\,
\bigg]^{-1}\;, \nonumber\\[3mm]
R^{-1}_t &=& \bigg[\, 1\: +\: \bigg(\, \frac{\alpha_s}{3\pi}\, 
e^{-i\phi_{\tilde{g}}}\: 
+\: \frac{|h_t|^2}{32\pi^2}\,
e^{-i\phi_{A_U}}\,\bigg)\,\tan\beta\,\bigg]^{-1}\;. 
\end{eqnarray}
Then,  from~(\ref{express}), it is easy to  see  that the off-diagonal
elements of  $\Vckm^\dagger \Ro^{-1} \Vckm$  increase if $\phi_{A_U},\
\phi_{\tilde{g}}   =   \pm  \pi$   and so   the    effective couplings
$g^{L,R}_{H_i\bar{d}    d'}$,  thereby   giving    rise   to  enhanced
predictions.   This   is a very   generic feature  which  is reflected
in~Fig.~\ref{fig2} and, as we will see in Section 4.2, also holds true
for our numerical estimates of $B$-meson FCNC observables.

\begin{figure}[p]
\psfig{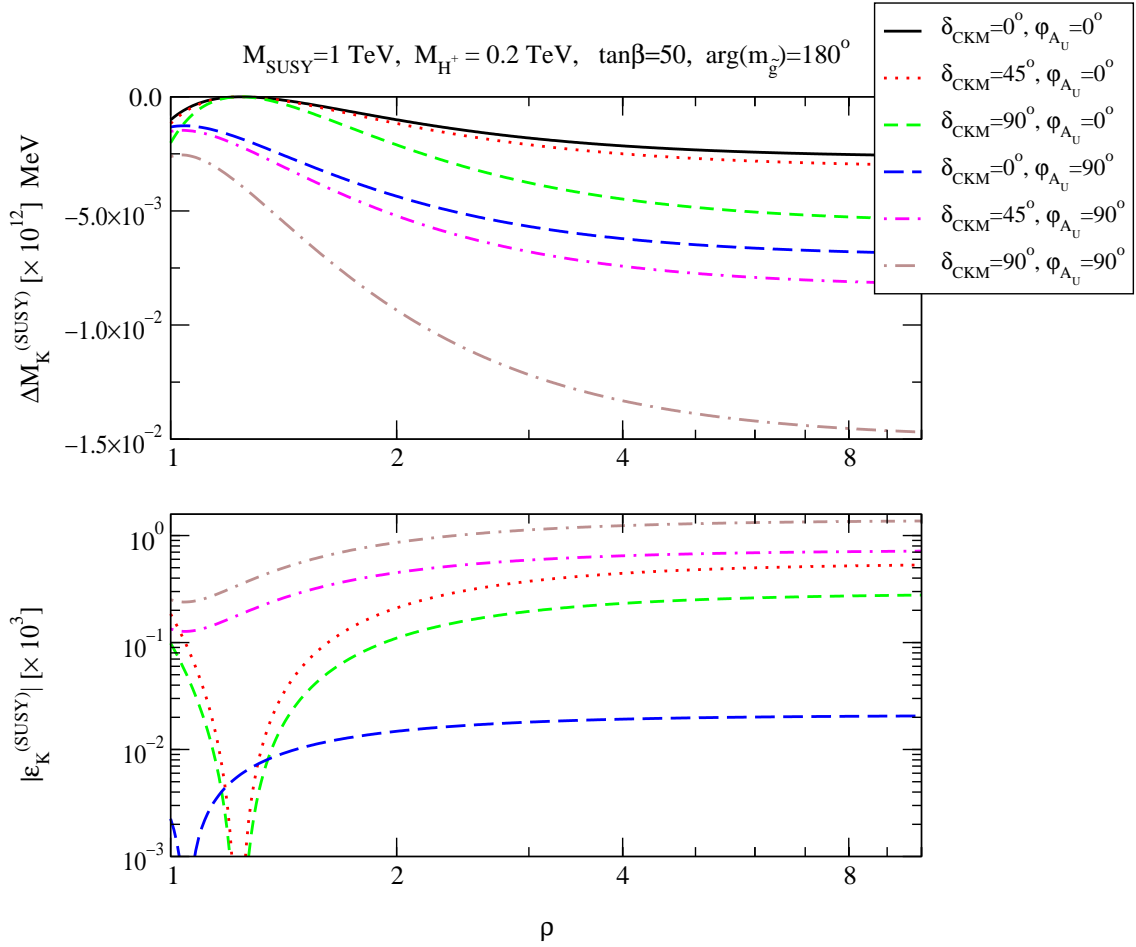}
\caption{{\it SUSY  Higgs-DP contributions to $\epsilon_K$ and $\Delta
M_K$ given in units of $10^{-12}$  MeV and $10^{-3}$, respectively, as
functions of the hierarchy factor $\rho$,  for $M_{\rm SUSY} = 1$~TeV,
$M_{H^+}   =   0.2$~TeV,    $\tan\beta =    50$,   and  ${\rm arg}\,
m_{\tilde{g}} = 180^\circ$, where the values of $\delta_{\rm CKM}$ and
$\phi_{A_U}$ are varied discretely.}}\label{fig3}
\end{figure}

Neglecting the small Yukawa couplings of the first two generations and
making use of the unitarity of $\Vckm$, we find for $d\neq d'$
\begin{eqnarray}
  \label{VRVapprox}
(\Vckm^\dagger \Ro^{-1} \Vckm)_{dd'}\ \propto\
V_{td}^*\, V_{td'}\, \Bigg[\, \frac{\alpha_s}{3\pi}\,e^{-i\phi_{\tilde{g}}}\,
\Bigg(\,\frac{1}{\rho^2}\: -\: 1\,\Bigg)\ -\ 
\frac{|h_t|^2}{32\pi^2}\,e^{-i\phi_{A_U}}\, \Bigg]\, \tan\beta\;. 
\end{eqnarray}
If $\rho=1$, the dominant FCNC  effect originates from the second term
in  the    square  brackets of~(\ref{VRVapprox}),   provided   $|1   +
\frac{\alpha_s}{3\pi}\, e^{-i\phi_{\tilde{g}}} \tan\beta | \gg 5\times
10^{-4}\,   \tan\beta$ (see also  footnote~2).  If  $\rho \gg 1$, then
gluino  corrections become dominant;  they  are  larger  by a   factor
$\frac{a_s}{3\pi}/\frac{1}{32\pi^2}\simeq  3.6$.  However, between the
low and high $\rho$-regime, there is an  intermediate value of $\rho$,
where $(\Vckm^\dagger \Ro^{-1} \Vckm)_{dd'}$  does exactly  vanish for
$d \neq d'$, and so the effective couplings $g^{L,R}_{H_i\bar{d} d'}$.
In this  case, one has $R_u^{-1}=R_c^{-1}=R_t^{-1}$ in (\ref{express}),
implying that $\Ro$ is  proportional to the unity  matrix. Then, it is
$(\Vckm^\dagger   \Ro^{-1}    \Vckm)_{dd'} = 0$,  as   a   result of a
GIM-cancellation mechanism due to the unitarity of  the CKM matrix. We
call  such a point in the  parameter space  {\em GIM operative point}.
The   $\rho$ value, for  which the  GIM-cancellation mechanism becomes
fully    operative,      may           easily      be       determined
from~(\ref{VRVapprox}),~i.e.\
\begin{eqnarray}
  \label{rgim}
\rho^2_{\rm GIM}\ =\ \bigg(\, 1\: +\: \frac{3|h_t|^2}{32\pi\alpha_s}\,
 e^{i(\phi_{\tilde{g}}-\phi_{A_U})}\, \bigg)^{-1} \;. 
\end{eqnarray}
For   the  MSSM  parameter  space  under   study,  there is  always  a
GIM-operative value for the hierarchy factor $\rho$, iff $\phi_{A_U} -
\phi_{\tilde{g}}   =   0$   or  $\pm   \pi$.    For    $\phi_{A_U}   -
\phi_{\tilde{g}} = 0$, we have  $\rho_{\rm  GIM} <  1$, whereas it  is
$\rho_{\rm GIM}  > 1$, for $\phi_{A_U} -  \phi_{\tilde{g}} = \pi$.  In
fact, the second  case is realized  in  Fig.~\ref{fig3} for $\rho_{\rm
GIM}    \approx    1.22$,   where    $\Delta     M_K^{\rm  SUSY}$  and
$|\epsilon_K^{\rm SUSY}|$ vanish independently of the value of the CKM
phase.     Here, we  should emphasize  the    fact that  the  value of
$\rho_{\rm   GIM}$ does  not depend    on the  FCNC  observable  under
consideration and is in  excellent agreement  with the one  determined
by~(\ref{rgim}).  Because of the  above flavour-universal  property of
$\rho_{\rm GIM}$, one  may even face  the very unusual possibility  of
discovering SUSY at high-energy colliders, without accompanying such a
discovery with any new-physics signal in low-energy $K$- and $B$-meson
FCNC observables.

It is   now instructive to  gauge  the relative size of  the different
DP-induced Wilson coefficients in~(\ref{dpkaon}).  For simplicity, let
us take $\rho=1$.  Then, each individual DP-induced Wilson coefficient
in~(\ref{dpkaon}) may be approximately given by
\begin{eqnarray}
  \label{sums}
\sum_{i=1}^3\,(g^L_{H_i\bar{s}d})^2 &\approx&  
(\chi_{\rm FC}^{(t)})^2\, (V_{ts}^*V_{td})^2\ \sum_{i=1}^3
\, \frac{O_{1i}^2 -O_{3i}^2+2\, i\, O_{1i} O_{3i}}{M_{H_i}^2} \ , 
\nonumber \\[3mm]
\sum_{i=1}^3\,(g^R_{H_i\bar{s}d})^2 &\approx&  
(\chi_{\rm FC}^{(t)*})^2\, (V_{ts}^*V_{td})^2\, \sum_{i=1}^3
\, \frac{O_{1i}^2 -O_{3i}^2-2\, i\, O_{1i} O_{3i} }{M_{H_i}^2} \ , 
\nonumber \\[3mm]
\sum_{i=1}^3\, (g^L_{H_i\bar{s}d}\, g^R_{H_i\bar{s}d}) &\approx&  
|\chi_{\rm FC}^{{(t)}}|^2\, (V_{ts}^*V_{td})^2\, \sum_{i=1}^3
\, \frac{O_{1i}^2 +O_{3i}^2 }{M_{H_i}^2} \ , 
\end{eqnarray}
where $\chi_{\rm FC}^{{(t)}}$ is the  $t$-quark dependent entry of the
diagonal  matrix  $\chi_{\rm  FC}$   defined  in~(\ref{chiFC}).    For
$M_{H^+} \stackrel{>}{{}_\sim} 180$~GeV, CP-violation and Higgs-mixing
effects start to decouple  from the lightest $H_1$-sector~\cite{CEPW}.
Moreover,  in   the region $\tan\beta \stackrel{>}{{}_\sim}   40$, the
$\Phi_2$-component in the $H_2$-  and $H_3$-boson mass-eigenstates  is
suppressed.  As a consequence of the latter, we obtain
\begin{eqnarray}
  \label{sum1}
\sum_{i=1}^3\ \frac{O_{1i}^2\: -\: O_{3i}^2}{M_{H_i}^2} &=&
\Big(\, O^2_{11}\: -\: O^2_{31}\,\Big)\, 
\Bigg(\,\frac{1}{M^2_{H_1}}\: -\: \frac{1}{M^2_{H_{23}}}\,\Bigg)\ +\
{\cal O}\Bigg(\,\frac{M^2_{H_2} - M^2_{H_3}}{M^2_{H_{23}}}\,\Bigg)\,,\\[3mm]
  \label{sum2}
\sum_{i=1}^3\ \frac{O_{1i}\, O_{3i}}{M_{H_i}^2} &=&
O_{11}\,O_{31}\, \Bigg(\,\frac{1}{M^2_{H_1}}\: -\:
\frac{1}{M^2_{H_{23}}}\,\Bigg)\ +\ 
{\cal O}\Bigg(\,\frac{M^2_{H_2} - M^2_{H_3}}{M^2_{H_{23}}}\,\Bigg)\,, 
\end{eqnarray}
where $M^2_{H_{23}} = \frac{1}{2}\, ( M^2_{H_2} + M^2_{H_3} )$ and the
orthogonality of the $O$ matrix has  been used.  Since it is $O_{11},\
O_{31} \ll 1$ in the kinematic region of our interest, then on account
of~(\ref{sum1})   and~(\ref{sum2})   and   for  maximal    CKM   phase
$\delta_{\rm CKM} = 90^\circ$, the   dominant contribution to  $\Delta
M_K^{\rm  SUSY}$ and  $\epsilon_K^{\rm SUSY}$ comes   from the last DP
expression in~(\ref{sums}), namely from the Wilson coefficient $C^{\rm
LR\, (DP)}_2$ in~(\ref{dpkaon}),  despite  the additional  suppression
factor $m_d/m_s \approx 1/10$ with respect  to $C^{\rm SLL\, (DP)}_1$.
If $\delta_{\rm CKM}= 0$, $\Delta  M_K^{\rm SUSY}$ still receives  its
largest      contribution   from    $C^{\rm     LR\, (DP)}_2$,   while
$|\epsilon_K^{\rm SUSY}|$  is dominated  by  the  first DP  expression
in~(\ref{sums}), i.e.\  from  $C^{\rm SLL\,  (DP)}_1$; the   second DP
expression in~(\ref{sums})  is   very suppressed with respect  to  the
first  one      by  two powers  of   the    ratio  $m_d/m_s$.   {}From
Fig.~(\ref{fig2}), we  also see  that  in addition   to the  CKM phase
$\delta_{\rm  CKM}$, the soft  SUSY-breaking  CP-phases, such as ${\rm
arg}\,(m_{\tilde{g}})$ and ${\rm arg}\, (A_U)$, may  also give rise by
themselves  to   enhancements   of  $|\Delta  M_K^{\rm    SUSY}|$  and
$|\epsilon_K^{\rm SUSY}|$  even    up to   one  order  of   magnitude.
Analogous remarks and  observations  also hold true for  the $B$-meson
FCNC observables which are to be discussed in the next section.

Finally, we  should comment on the  fact that the 2HDM contribution by
itself due to $C^{\rm   LR\,(2HDM)}_2$ in~(\ref{2HDM}) can  only  give
rise   to the undetectably  small  numerical values, $|\Delta M_K^{\rm
2HDM}|=5\times  10^{-17}$ MeV and  $|\epsilon_K^{\rm 2HDM}|  = 5\times
10^{-6}$ (indicated  by   an  arrow   in  the  Fig.~\ref{fig2}),   for
$\delta_{\rm CKM}=90^\circ$.

\subsection{$\Delta M_{B_q}$, $B_q\to \ell^+\ell^-$ and associated leptonic 
  CP asymmetries}

In this  section, we will present numerical  estimates for a number of
$B$-meson FCNC observables, such     as the mass   difference  $\Delta
M_{B_q}$, the branching  ratio for  $B_q\to  \ell^+\ell^-$ and the  CP
asymmetries  associated with  the  $B$-meson   leptonic decays.    The
current  experimental  status  of       these  observables   is     as
follows~\cite{PDG}:
\begin{eqnarray}
  \label{expBd}
\Delta M_{B_d} & = & 0.489\: \pm\: 0.008\ {\rm ps}^{-1}\,,\\
  \label{expBs}
\Delta M_{B_s} & > & 13.1\ {\rm ps}^{-1}\,,\\
  \label{expBsmu}
{\cal B}( B_s \to \mu^+\mu^- ) & < & 2.0\times 10^{-6}
\end{eqnarray}
and~\cite{Ligeti}
\begin{equation}
  \label{expBdtau}
{\cal B}( B_d \to \tau^+ \tau^- ) \ <\ 0.015\, .
\end{equation}
Future experiments at an upgraded  phase of the Tevatron collider  may
reach higher sensitivity to ${\cal B}( B_s \to \mu^+\mu^- )$ up to the
$10^{-8}$-level~\cite{Anikeev,ADKT,Dedes1}.

\begin{figure}[p]
\psfig{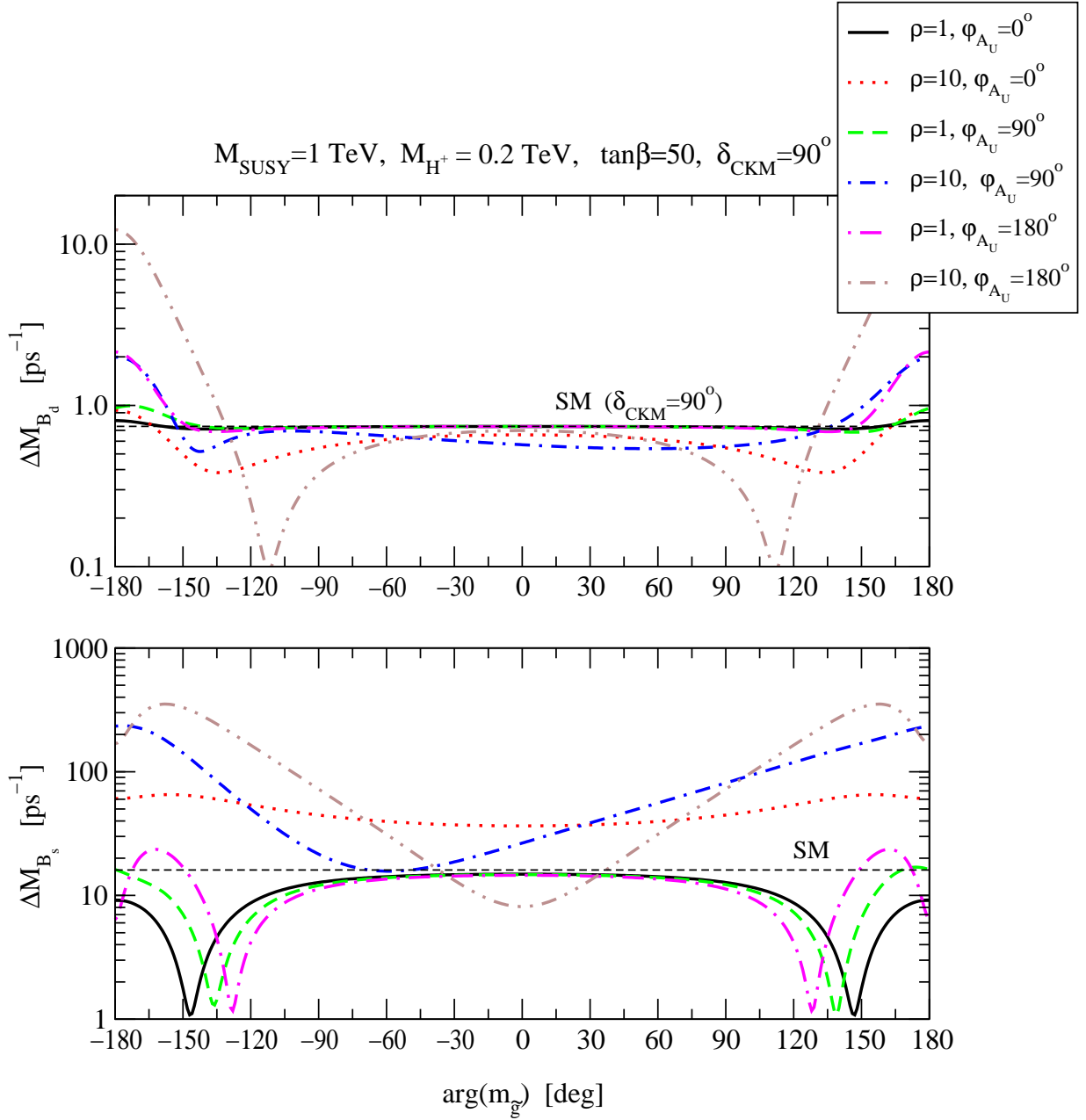}
\caption{{\it SM and SUSY Higgs-DP contributions to $\Delta M_{B_d}$ and
$\Delta M_{B_s}$ as functions of the gluino phase ${\rm arg}\,
(m_{\tilde{g}})$, for $M_{\rm SUSY} = 1$~TeV, $M_{H^+} = 0.2$~TeV,
$\tan\beta = 50$ and $\delta_{\rm CKM} = 90^\circ$, where the
hierarchy factor $\rho$ and $\phi_{A_U}$ are varied independently as
shown above. The SM contributions alone for $\delta_{\rm CKM} = 90^\circ$
are displayed by horizontal dashed lines.}}\label{fig4}
\end{figure}

\begin{figure}[p]
\psfig{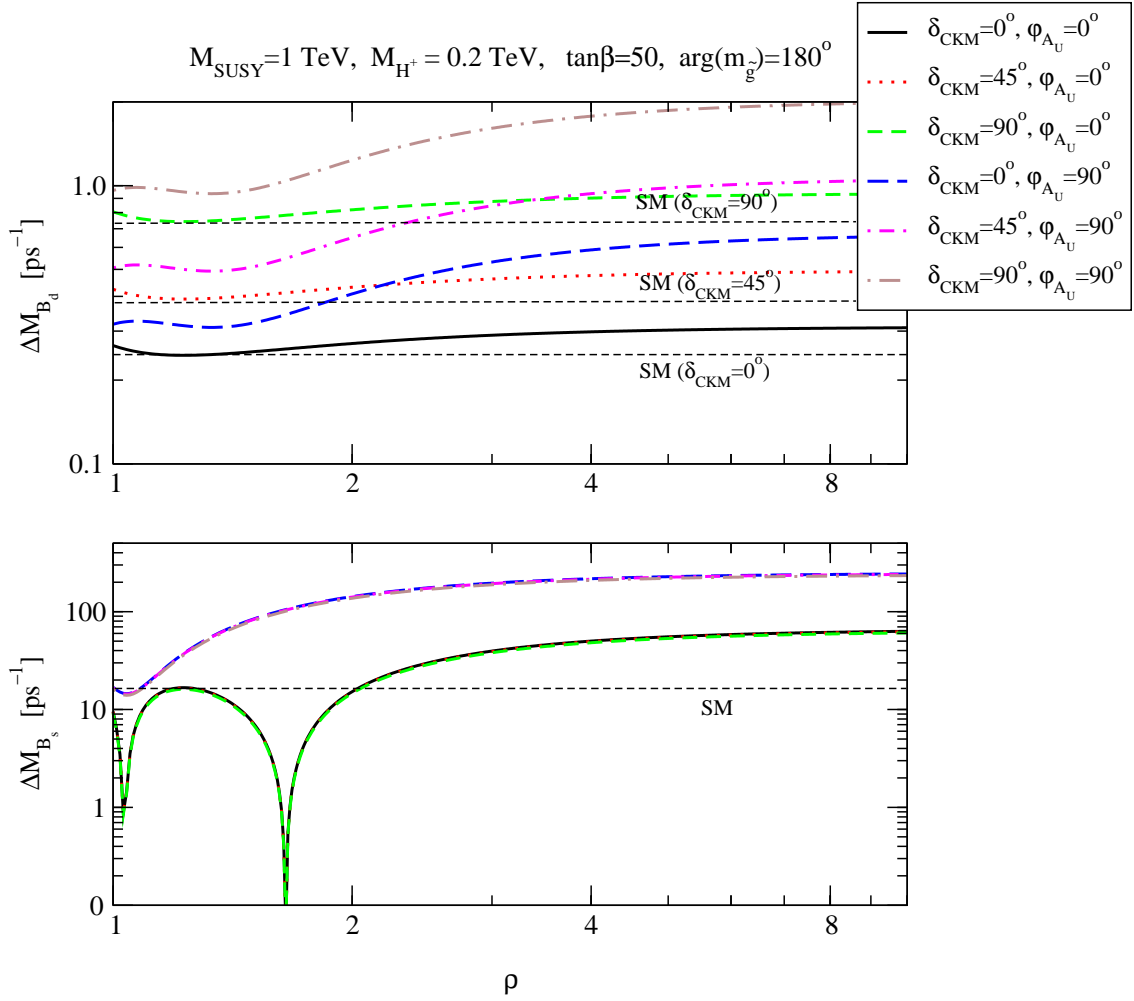}
\caption{{\it SM and SUSY Higgs-DP contributions to $\Delta M_{B_d}$
and $\Delta M_{B_s}$ versus the hierarchy factor $\rho$, for $M_{\rm
SUSY} = 1$~TeV, $M_{H^+} = 0.2$~TeV, $\tan\beta = 50$, and ${\rm
arg}\, m_{\tilde{g}} = 180^\circ$, where $\delta_{\rm CKM}$ and
$\phi_{A_U}$ obtain discrete values as shown above. Also shown are the
SM effects alone for different choices of the CKM phase $\delta_{\rm
CKM}$ (horizontal dashed lines).}}\label{fig5}
\end{figure}

Let us  start our  discussion  by numerically analyzing the  $B$-meson
mass differences $\Delta M_{B_d}$ and $\Delta M_{B_s}$. As in the case
of  the $K$-meson observables, we use  the same  input values as those
shown in Fig.~\ref{fig1},  i.e.\  $M_{\rm SUSY}  = 1$~TeV,  $M_{H^+} =
0.2$~TeV  and $\tan\beta  =  50$.  Then,  Fig.~\ref{fig4} displays the
combined, SM and   Higgs-DP,  contributions to   $\Delta  M_{B_d}$ and
$\Delta M_{B_s}$  as functions   of the   gluino phase  ${\rm   arg}\,
(m_{\tilde{g}})$, for  $\delta_{\rm CKM}   = 90^\circ$  and  different
choices of hierarchy factor $\rho$ and $\phi_{A_U}$.  Note that the SM
contributions alone for $\delta_{\rm CKM} = 90^\circ$ are displayed by
horizontal dashed  lines.  Even though the  SM predictions for $\Delta
M_{B_{d,s}}$ may  adequately describe by  themselves the  experimental
values in~(\ref{expBd})  and (\ref{expBs}), they cannot yet decisively
exclude  possible new-physics   contributions   due to  the   inherent
uncertainties in the calculation  of hadronic matrix elements such  as
those  induced  by   SUSY    Higgs-mediated FCNC     interactions.  In
particular, we observe in Fig.~\ref{fig4} that SM and Higgs-DP effects
may   add constructively  or   destructively  to the mass  differences
$\Delta M_{B_{d,s}}$.   Similar features are found in~Fig.~\ref{fig5},
where the  SM and SUSY Higgs-DP contributions  to $\Delta M_{B_d}$ and
$\Delta M_{B_s}$ are plotted  versus the hierarchy factor  $\rho$, for
discrete  values  of $\delta_{\rm CKM}$ and  $\phi_{A_U}$.   In our SM
CKM-phase convention~\cite{PDG},   unlike   the  CKM matrix    element
$V_{ts}$,   the  matrix  element   $V_{td}$   is  very   sensitive  to
$\delta_{\rm  CKM}$  values.   As  a result,  the  SM  predictions for
$\Delta M_{B_d}$ strongly depend on the selected value of $\delta_{\rm
CKM}$, as can be seen from~Fig.~\ref{fig5}.

\begin{figure}[p]
\psfig{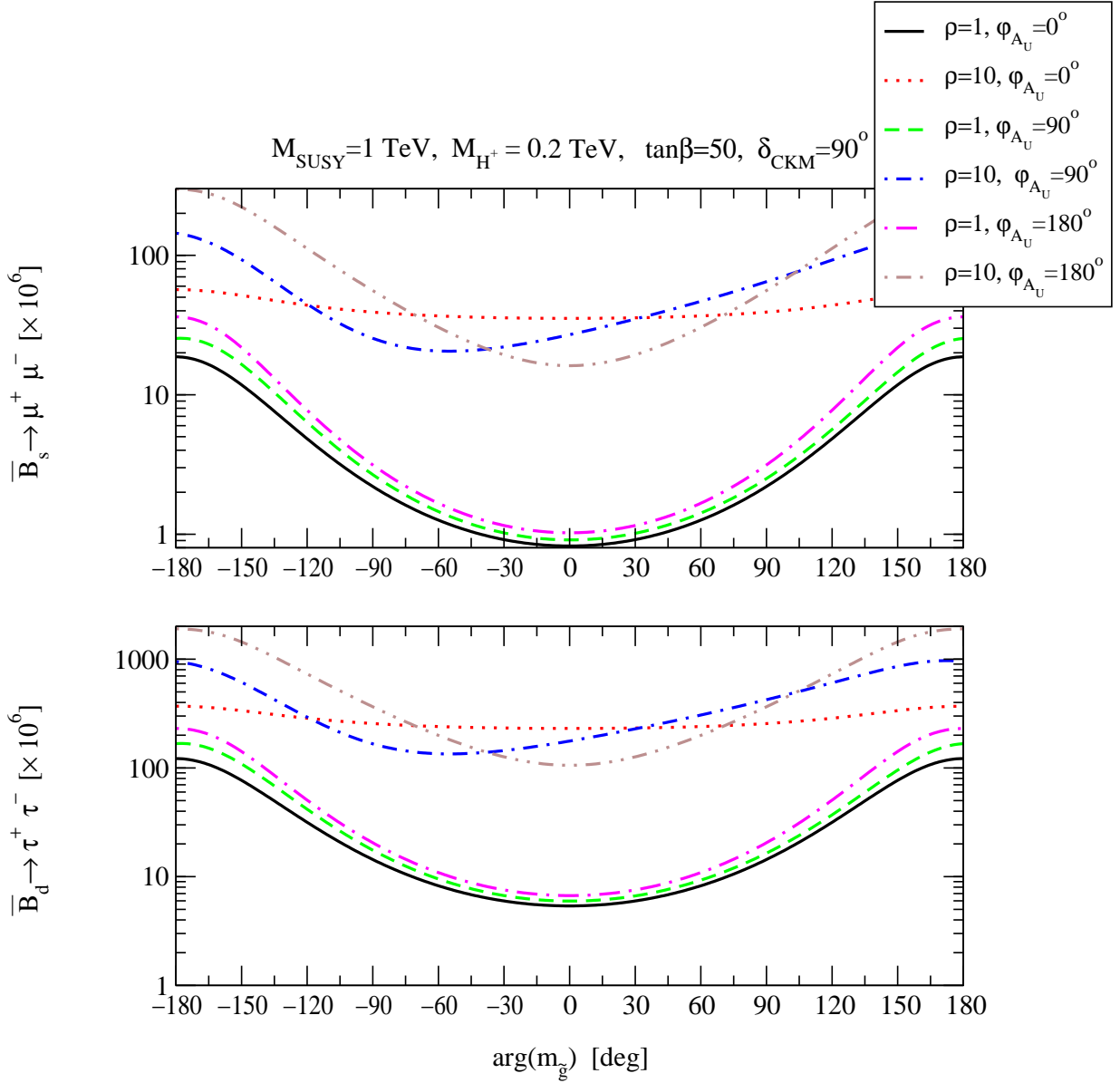}
\caption{{\it SUSY Higgs-penguin contributions to ${\cal B}
(\bar{B}^0_s \to \mu^+\mu^-)$ and ${\cal B} (\bar{B}^0_d \to
\tau^+\tau^-)$ versus the gluino phase ${\rm arg}\, (m_{\tilde{g}})$,
for $M_{\rm SUSY} = 1$~TeV, $M_{H^+} = 0.2$~TeV, $\tan\beta = 50$, and
$\delta_{\rm CKM} = 90^\circ$, where $\rho$ and $\phi_{A_U}$ are
varied discretely.}}\label{fig6}
\end{figure}

In Fig.~\ref{fig6}, we  exhibit   numerical values for the   branching
ratios  ${\cal  B}  (\bar{B}^0_s  \to    \mu^+\mu^-)$  and  ${\cal  B}
(\bar{B}^0_d   \to \tau^+\tau^-)$ as  functions   for the gluino phase
${\rm arg}\, (m_{\tilde{g}})$, for  $M_{\rm SUSY} = 1$~TeV, $M_{H^+} =
0.2$~TeV, $\tan\beta  = 50$, and  $\delta_{\rm CKM} = 90^\circ$, where
$\rho$ and $\phi_{A_U}$   are varied discretely.   Since the branching
ratios are driven by   Higgs-penguin effects in   the region of  large
$\tan\beta$, predictions for other inputs of $\tan\beta$ and $M_{H^+}$
may be easily estimated by rescaling  the numerical values by a factor
$x_{\cal O} =  (\tan\beta/50 )^6\times (0.2~{\rm TeV}/M_{H^+})^4$, for
$\tan\beta \stackrel{>}{{}_\sim}   40$.   Thus,     confronting    the
predictions  for   ${\cal  B}  (\bar{B}^0_s  \to    \mu^+\mu^-)$  with
experiment   data   in~(\ref{expBsmu}),    combined   bounds  on   the
$\tan\beta$-$M_{H^+}$ plane may be obtained   for a given set of  soft
SUSY-breaking parameters. As we see in~Fig.~\ref{fig6}, these combined
bounds become even  more  restrictive for  large gluino phases,  ${\rm
arg}\, m_{\tilde{g}}   \stackrel{>}{{}_\sim} 90^\circ$,  in  agreement
with our discussions in Section~4.1.

\begin{figure}[p]
\psfig{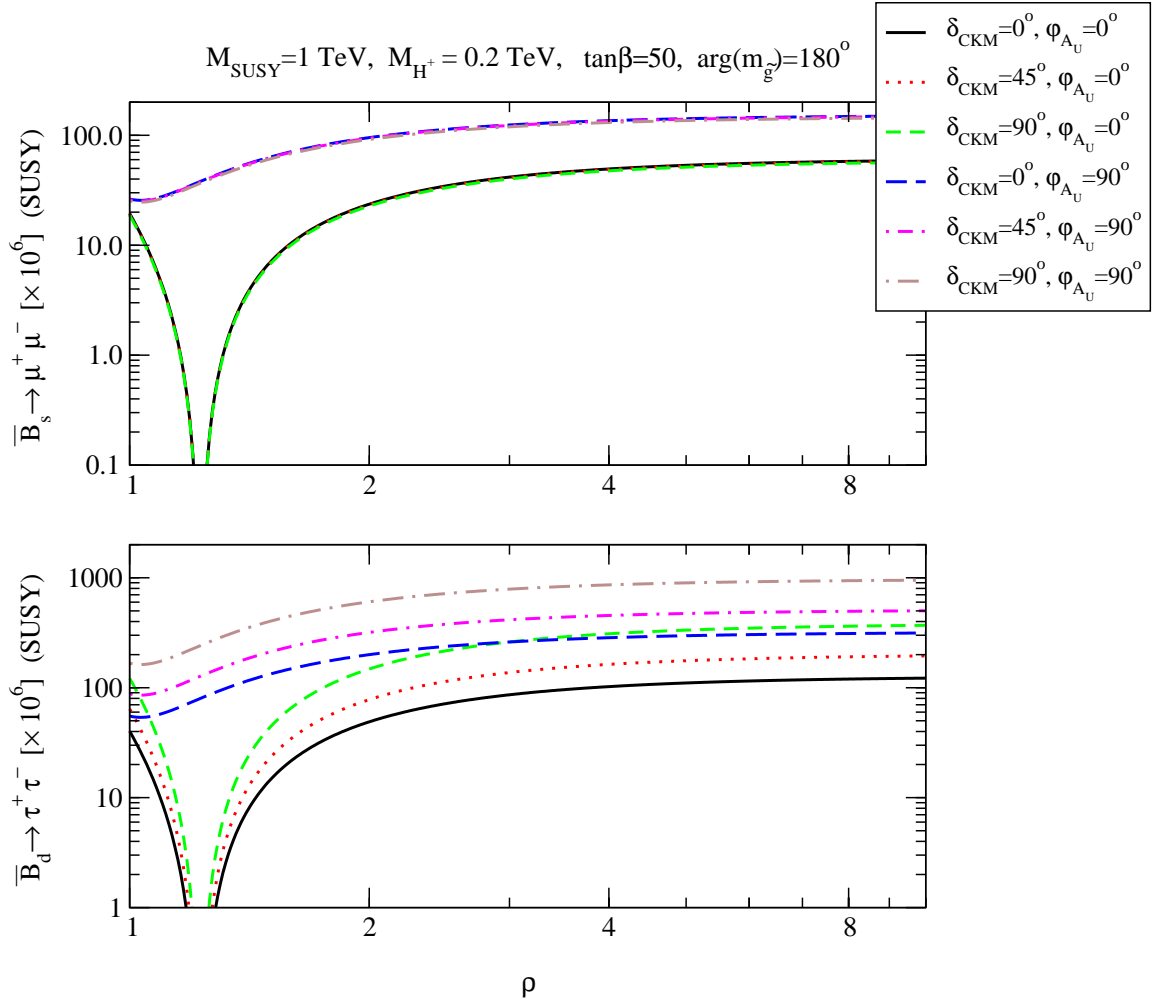}
\caption{{\it SUSY Higgs-penguin contributions to ${\cal B}
(\bar{B}^0_s \to \mu^+\mu^-)$ and ${\cal B} (\bar{B}^0_d \to
\tau^+\tau^-)$ as functions of the hierarchy factor $\rho$, for
$M_{\rm SUSY} = 1$~TeV, $M_{H^+} = 0.2$~TeV, $\tan\beta = 50$, and
${\rm arg}\, m_{\tilde{g}} = 180^\circ$, where $\delta_{\rm CKM}$ and
$\phi_{A_U}$ take discrete values.}}\label{fig7}
\end{figure}

However, there  is an additional factor  that may crucially affect our
predictions for  the branching ratios of   the decays $\bar{B}^0_s \to
\mu^+\mu^-$ and $\bar{B}^0_d  \to \tau^+\tau^-$,  namely the hierarchy
parameter $\rho$.  As we show in Fig.~\ref{fig7}, even for the extreme
choice  of a gluino  phase ${\rm  arg}\,  m_{\tilde{g}}  = 180^\circ$,
${\cal B} (\bar{B}^0_d \to \mu^+\mu^-)$ and ${\cal B} (\bar{B}^0_s \to
\mu^+\mu^-)$ can get very suppressed for a specific value of $\rho$ in
certain soft SUSY-breaking scenarios that  can realize a GIM-operative
point in their  parameter space.  As we  detailed in Section~4.1, this
phenomenon occurs for the universal value of  $\rho = \rho_{\rm GIM} =
1.22$, when $\phi_{A_U} - \phi_{\tilde{g}}  = \pm 180^\circ$.  As  can
be seen in Fig.~\ref{fig7}, the  predicted values for $\bar{B}^0_s \to
\mu^+\mu^-$ and $\bar{B}^0_d   \to \tau^+\tau^-$, where $\phi_{A_U}  =
0^\circ$, confirm the above observation.

As we have  already  mentioned, the  observables  $\Delta M_{B_s}$ and
${\cal B}(\bar{B}_s^0\to \mu^+  \mu^-)$  exhibit a different   scaling
behaviour with  respect   to $M_{H^+}$  and $\tan\beta$,  through  the
scaling   factor $x_{\cal O}$  in~(\ref{scale}).   Once  the above two
kinematic   parameters   are fixed to   some   input  values,  the two
observables   $\Delta  M_{B_s}$   and  ${\cal B}(\bar{B}_s^0\to  \mu^+
\mu^-)$  are  then rather   correlated   to each  other, since   their
dependences on the  $\mu$-parameter and $A_U$  are very similar in the
minimal flavour-violating case $\rho =1$~\cite{Buras}.  As one can see
from Figs.~\ref{fig4}--\ref{fig7}, our  numerical analysis agrees well
with the  above result for $\rho =1$.   However,  we also observe that
the correlation is practically lost for $\rho > 1$, e.g.\ close to the
GIM-operative points, and/or by the inclusion of CP-violating effects,
as $\Delta M_{B_s}$  and  ${\cal B}(\bar{B}_s^0\to \mu^+ \mu^-)$  have
different dependences on the soft CP-odd phases.

\begin{figure}[p]
\psfig{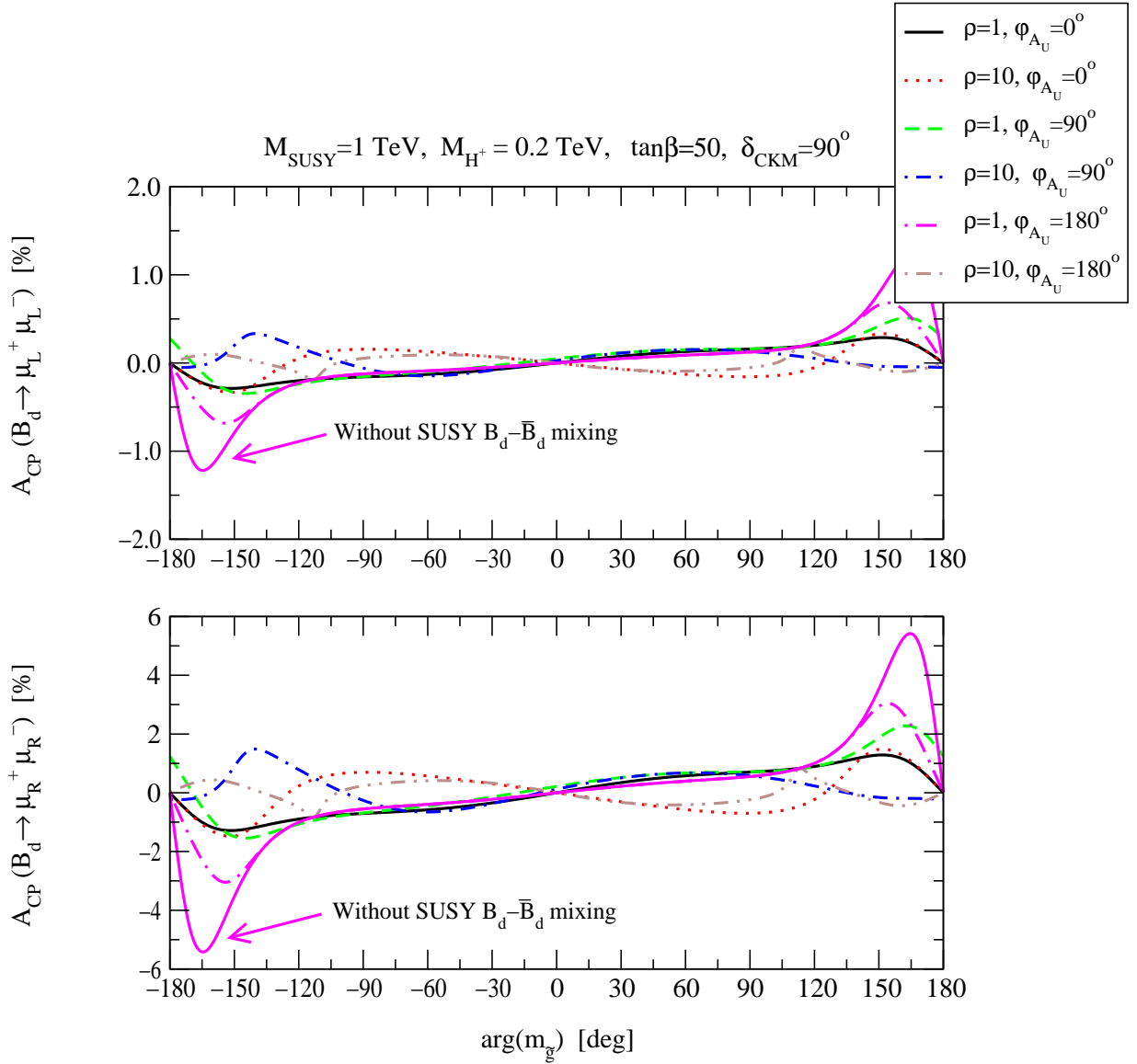}
\caption{{\it Numerical values for the CP asymmetries ${\cal A}_{\rm
CP}^{(B^0_d \to \mu^+_L\mu^-_L )}$ and ${\cal A}_{\rm
CP}^{(B^0_d \to \mu^+_R\mu^-_R )}$ as functions of the gluino
phase ${\rm arg}\, (m_{\tilde{g}})$, for $M_{\rm SUSY} = 1$~TeV,
$M_{H^+} = 0.2$~TeV, $\tan\beta = 50$, and $\delta_{\rm CKM} =
90^\circ$, where $\rho$ and $\phi_{A_U}$ are varied
discretely. Also shown is the prediction for the CP asymmetries
without including $B^0_d$-$\bar{B}^0_d$ mixing.}}\label{fig8}
\end{figure}

Let us now investigate the size of the  CP asymmetries in the leptonic
$B_d$-meson decays  in  the  CP-violating MSSM;  the  corresponding CP
asymmetries for  the $B_s$ meson are  experimentally constrained to be
rather small, less than   5\% (see also discussion  after~(\ref{lq})).
In Fig.~\ref{fig8}, we display numerical values for the CP asymmetries
${\cal A}_{\rm CP}^{(B^0_d \to \mu^+_L\mu^-_L  )}$ and ${\cal  A}_{\rm
CP}^{(B^0_d  \to \mu^+_R\mu^-_R )}$  as functions of  the gluino phase
${\rm arg}\, (m_{\tilde{g}})$, for $M_{\rm SUSY}  = 1$~TeV, $M_{H^+} =
0.2$~TeV,   $\tan\beta = 50$, and  $\delta_{\rm  CKM} = 90^\circ$.  As
usual, we independently vary the parameters $\rho$ and $\phi_{A_U}$ to
take on the discrete values $\rho = 1,\ 10$ and $\phi_{A_U} = 0^\circ,
90^\circ$ and $180^\circ$.  We  find that if the $B^0_d$-$\bar{B}^0_d$
mixing is consistently taken into account, the  typical size of ${\cal
A}_{\rm   CP}^{(B^0_d  \to \mu^+_L\mu^-_L   )}$   and ${\cal   A}_{\rm
CP}^{(B^0_d  \to  \mu^+_R\mu^-_R )}$  does not  exceed 0.7\%  and 3\%,
respectively. If $B^0_d$-$\bar{B}^0_d$  mixing is not included, the CP
asymmetries  can reach slightly  higher  values up to  1.2\%  and 6\%,
respectively.  The  apparent   reason for  the  smallness  of   the CP
asymmetries is due to the occurrence of an approximate cancellation in
the  sum $C_S + C_P$ in~(\ref{lq})  at large $\tan\beta$,  as the muon
velocity is $\beta_\mu \approx 1$.

Having gained  some insight  from  the  above exercise,  one may  seek
alternative ways  to enhance  the di-muon asymmetries   ${\cal A}_{\rm
CP}^{(B_d^0\to \mu_{L,(R)}^+ \mu_{L,(R)}^-)}$.  To this end, the first
attempt would be to  suppress  the effect of the   $B_d^0-\bar{B}_d^0$
mixing  by considering smaller   $\tan\beta$ values, e.g.\  $\tan\beta
\stackrel{<}{{}_\sim} 10$. In this intermediate region of $\tan\beta$,
the above cancellation in  the sum $C_S + C_P$  does not occur due  to
non-trivial CP-violating  Higgs-mixing    effects  and  so    the   CP
asymmetries  ${\cal  A}_{\rm       CP}^{(B_d^0\to        \mu_{L,(R)}^+
\mu_{L,(R)}^-)}$ can be significantly increased. To get an idea of the
magnitude of   the  CP asymmetries   in this  case,   we  consider the
so-called  CPX   scenario  introduced   in~\cite{CPX}   to    maximize
CP-violating effects in the    lightest Higgs sector of an   effective
MSSM. In the CPX scenario,  the $\mu$-parameter and the soft trilinear
Yukawa coupling $A_U$ are set by the relations:  $\mu= 4 M_{\rm SUSY}$
and $A_U=  2  M_{\rm SUSY}$.  Thus,  for   $M_{H^+}=0.15$ TeV, $M_{\rm
SUSY}=1$  TeV, $\tan\beta=7$,    $\rho=10$, $\phi_{A_U}=45^\circ$  and
$\phi_{\tilde{g}}=0^\circ$,  we  find that  CP-violating Higgs-penguin
effects can give rise to the CP asymmetries:
\begin{equation}
  \label{CPX1}
{\cal A}_{\rm CP}^{(B_d^0\to \mu_L^+ \mu_L^-)}\ \approx\ -9~\%\,,\qquad 
{\cal A}_{\rm CP}^{(B_d^0\to \mu_R^+ \mu_R^-)}\ \approx\ -37~\%
\end{equation} 
where ${\cal B}(B_d\to \mu^+  \mu^-)=3.6 \times 10^{-10}$, which is an
order  of  magnitude  larger  than   the SM  prediction~\cite{Dedes2}.
Another variant   of   the CPX scenario  of   equally phenomenological
importance utilizes the    parameters: $\tan\beta=10$, $\rho  =   10$,
$\phi_{A_U}=45^\circ$  and   $\phi_{\tilde{g}}=90^\circ$, with $M_{\rm
SUSY} = 1$~TeV. In this case, we obtain
\begin{equation}
  \label{CPX2}
{\cal A}_{\rm CP}^{(B_d^0\to \mu_L^+ \mu_L^-)}\ \approx\ 11~\%\,,\qquad
{\cal A}_{\rm CP}^{(B_d^0\to \mu_R^+ \mu_R^-)}\ \approx\ 43~\% 
\end{equation}
and ${\cal B}(B_d\to \mu^+  \mu^-)=4.1  \times 10^{-9}$, which is  two
orders  of  magnitude  above the SM   prediction.   The above scenario
appears to pass  all   the experimental constraints,   including those
deduced      from      LEP2     analyses      of     direct      Higgs
searches~\cite{CPX,Hreview}.   Most interestingly,      a     possible
observation of a non-zero CP asymmetry in the leptonic di-muon channel
will constitute  the harbinger for new  physics at  $B$ factories.  At
this stage, it is important to comment on  the fact that the numerical
values  stated  in~(\ref{CPX1}) and~(\ref{CPX2})  should  be viewed as
crude estimates, since they are obtained entirely on  the basis of our
resummed  FCNC effective Lagrangian~(\ref{master}) at the intermediate
$\tan\beta$ regime.  However, in this region of $\tan\beta$, we expect
additional  one-loop  effects  to  start  getting relevant,   such  as
supersymmetric $Z$-penguin and box  diagrams.  Even though our initial
estimates  given above appear  to yield  rather encouraging results, a
complete study  of the  leptonic $B$-meson  branching  ratios  and the
respective CP asymmetries for    all values of $\tan\beta$   would  be
preferable.

\begin{figure}[p]
\psfig{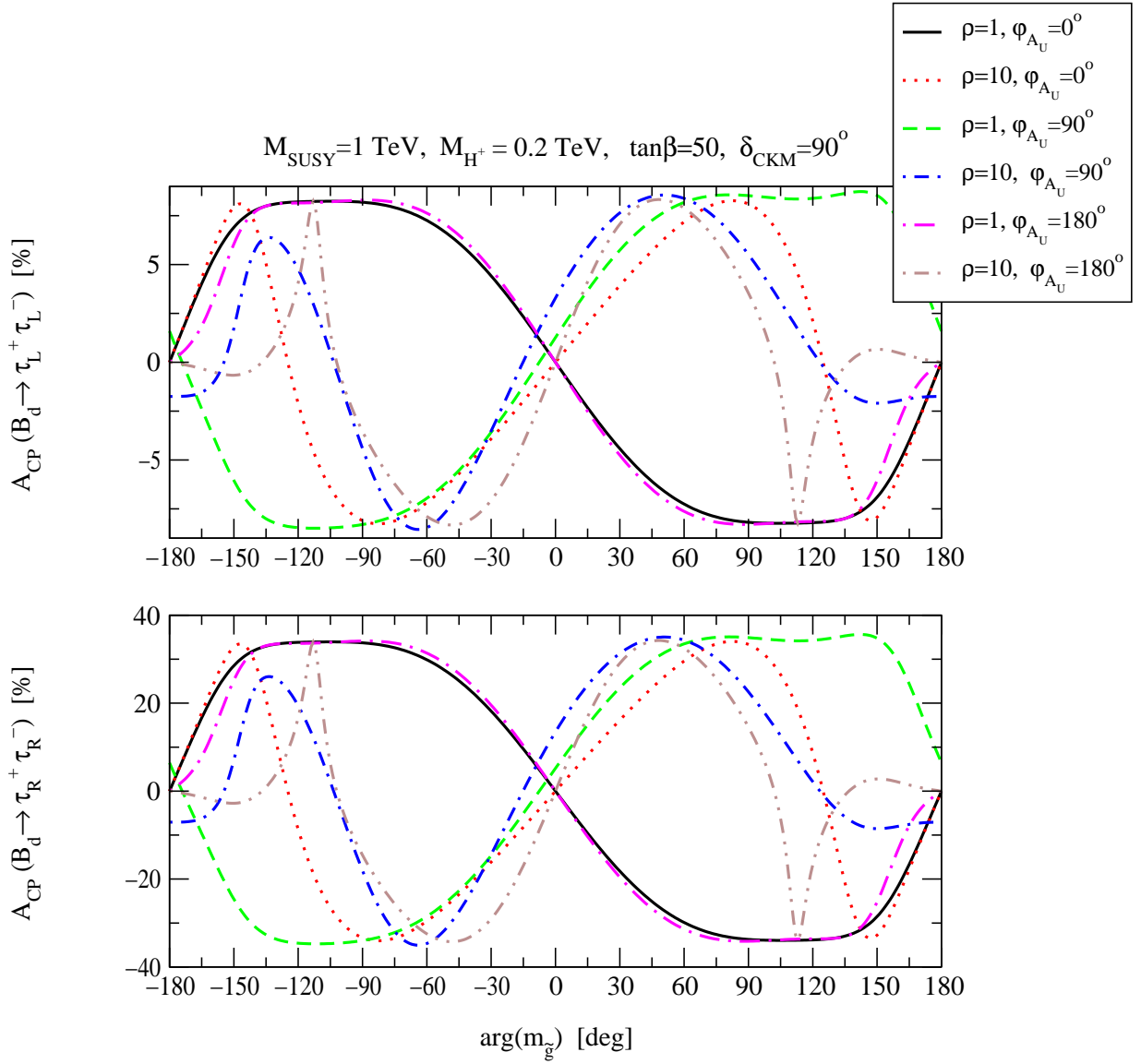}
\caption{{\it Numerical estimates of the CP asymmetries ${\cal A}_{\rm
CP}^{(B^0_d \to \tau^+_L\tau^-_L )}$ and ${\cal A}_{\rm
CP}^{(B^0_d \to \tau^+_R\tau^-_R )}$ versus the gluino phase
${\rm arg}\, (m_{\tilde{g}})$, for $M_{\rm SUSY} = 1$~TeV, $M_{H^+} =
0.2$~TeV, $\tan\beta = 50$, and $\delta_{\rm CKM} = 90^\circ$, where
$\rho$ and $\phi_{A_U}$ take discrete values as shown
above.}}\label{fig9}
\end{figure}

In  the  case of $\tau$-lepton  CP  asymmetries, the velocities of the
decayed  $\tau$-leptons $\beta_\tau$ is  roughly 0.5, so one naturally
gets an appreciably higher value for  the expression $\beta_\tau\, C_S
+ C_P \approx 0.5 C_P$ in~(\ref{lq}).  As a  result, larger values for
the  $\tau$-lepton    CP    asymmetries   are  expected.  Indeed,   in
Fig.~\ref{fig9}, we display numerical  predictions for ${\cal  A}_{\rm
CP}^{(B^0_d \to   \tau^+_L\tau^-_L )}$ and  ${\cal A}_{\rm CP}^{(B^0_d
\to  \tau^+_R\tau^-_R    )}$ versus the    gluino  phase  ${\rm arg}\,
(m_{\tilde{g}})$,  for the same values  of the input  parameters as in
Fig.~\ref{fig8}. Then, the CP  asymmetries ${\cal A}_{\rm  CP}^{(B^0_d
\to  \tau^+_L\tau^-_L )}$  and      ${\cal A}_{\rm CP}^{(B^0_d     \to
\tau^+_R\tau^-_R )}$ can be as high as 9\% and 36\%, respectively.

We conclude this section with some general remarks. In addition to the
$K$- and $B$-meson observables we have been studying  here, there is a
large number of other FCNC observables which have  to be considered in
a combined  full-fledged analysis. For example, the  decay $B  \to X_s
\gamma$~\cite{Bsgamma1,Bsgamma2}   plays a  central r\^ole   in such a
global  analysis,  because   it will  enable    us to  delineate  more
accurately the CP-conserving/CP-violating soft SUSY-breaking parameter
space   favoured by  low-energy  FCNC observables.    In this context,
constraints on  CP-violating SUSY phases  from  the non-observation of
electron  and  neutron electric dipole moments   (EDMs) should also be
implemented.  Large CP gluino and stop phases,  as the ones considered
in our  analysis, would require either  very heavy squarks with masses
larger than 5-6~TeV or the existence of a cancellation mechanism among
the       different     one-,    two-       and    higher-loop     EDM
contributions~\cite{EDM1,EDM2,APEDM}.   Especially,  it has been shown
recently~\cite{APEDM} that if the first two  generation of squarks are
heavier  than about 3~TeV, the required  degree  of cancellations does
not exceed the 10\% level and hence large CP-violating gluino, gaugino
and third-generation phases are still allowed for  wide regions of the
MSSM parameter space.  Finally, in  the present numerical analysis, we
have concentrated on scenarios that minimally  depart from the minimal
flavour-violation assumption  through  the  presence of diagonal,  but
non-universal squark masses.  In the  most general case, however,  the
squark mass matrices and consequently the  resummation matrix $\Ro$ of
the radiative threshold effects  may not be diagonal.  Such low-energy
realizations with off-diagonal  soft squark-mass matrices can still be
treated  exactly  within the context  of  our resummed  FCNC effective
Lagrangian~(\ref{master}),  by appropriately  considering  non-trivial
quark-squark   CKM-like matrices, such  as  the  3-by-3 unitary matrix
$\ULQ$ in~(\ref{hd}).

\setcounter{equation}{0} 
\section{Conclusions}

We   have derived the  general  form  for the  effective Lagrangian of
Higgs-mediated   FCNC   interactions    to  $d$-type   quarks,   where
large-$\tan\beta$   radiative threshold  effects   have been  resummed
consistently (cf.~(\ref{master}) and (\ref{couplings})).  Our resummed
FCNC  effective Lagrangian is   free from  pathological singularities,
which  mainly  emanate  from    the top-quark   dominance   hypothesis
frequently adopted in    the literature, and has   been  appropriately
generalized  to  include  effects of  non-universality   in the squark
sector,  as  well  as   CP-violation  effects  originating   from  the
CKM-mixing  matrix  and  the  complex  soft  SUSY-breaking masses.  In
particular, our resummed effective Lagrangian can  be applied to study
Higgs-mediated    FCNC effects   in   more general  soft SUSY-breaking
scenarios, beyond  those that have  already been discussed  within the
restricted framework  of models with  minimal flavour violation. Also,
an approach to resumming  radiative threshold effects,  very analogous
to the one developed in Section~2, can straightforwardly be applied to
see-saw SUSY   models,   so as to   properly   describe Higgs-mediated
lepton-flavour-violating interactions.
 
Within the context of   generic soft SUSY-breaking scenarios, we  have
analyzed a  number of $K$- and $B$-meson  observables, such as $\Delta
M_{K,B}$, $\epsilon_K$,  $\epsilon'/\epsilon$,  ${\cal   B}(B_{s,d}\to
\ell^+\ell^-)$ and their associated leptonic  asymmetries~\cite{code},
which  are enhanced by  Higgs-boson FCNC interactions for large values
of $\tan\beta$.  We  have found that  the predictions crucially depend
on the   choice of soft CP-violating  phases  in a  given  set of soft
SUSY-breaking parameters.  For  example,  for  certain values of   the
gluino and stop phases, the predictions can reach  and even exceed the
current  experimental limits, whereas for other   values of the CP-odd
phases the  FCNC effects can  be reduced by  one or even two orders of
magnitude.    Most   remarkably,  we  have    been  able  to  identify
configurations in  the soft  SUSY-breaking   parameter space, such  as
$\rho_{\rm GIM}$, where a kind of a GIM-cancellation mechanism becomes
fully operative [cf.~(\ref{rgim})] and, as a result of the latter, all
Higgs-mediated, $\tan\beta$-enhanced  effects   on $K$- and  $B$-meson
FCNC observables are completely absent.

Based on our resummed effective Lagrangian, one may now carry over the
present analysis  to a   vast number  of   other $K$-  and   $B$-meson
observables.  Evidently, further  dedicated studies  need be performed
in this direction.  We expect the obtained predictions to affect other
low- and high-energy observables, such as measurements of electron and
neutron electric dipole  moments and Higgs-boson  searches, as well as
studies on cosmological electroweak baryogenesis and  dark matter.  It
would be very  interesting to determine  to which degree  the emerging
CP-violating  MSSM  framework  with  CP-mixed  Higgs bosons  mediating
$\tan\beta$-enhanced   interactions   to matter   could be potentially
responsible for all  the present and future CP-conserving/CP-violating
FCNC effects observed in nature.
 
\subsection*{Acknowledgements}
We wish to  thank Ulrich Nierste and Alexander  Kagan for illuminating
discussions,  and  Herbi  Dreiner   for  a  critical  reading  of  the
manuscript.  A.D.  acknowledges financial support from the CERN Theory
Division  and  the  Network  RTN European  Program  HPRN-CT-2000-00148
``Physics Across  the Present Energy  Frontier: Probing the  Origin of
Mass.'' A.P.   thanks the Fermilab  Theory Group for  warm hospitality
and support.

\newpage

\begin{thebibliography}{99}
\bibitem{comment} For example, gaugino-Higgsino mixing effects studied
  in~\cite{MTW}  and~\cite{IN} can be  straightforwardly included in our
  formalism.
\end{thebibliography}
\end{document}